\documentclass[11pt]{article}
\usepackage[margin=1in]{geometry}
\usepackage{amsmath}
\usepackage{amssymb}
\usepackage{graphicx}
\usepackage{chicago}
\usepackage{booktabs}
\usepackage{enumerate}
\usepackage{setspace}
\usepackage{bm}
\usepackage{dsfont}
\usepackage[update,prepend]{epstopdf}
\usepackage{multirow}
\usepackage{multicol}
\usepackage{dcolumn}
\usepackage{bbm}
\usepackage{color}
\usepackage[T1]{fontenc}
\usepackage{authblk}
\usepackage{ntheorem}
\usepackage{float}
\usepackage{mathrsfs}
\usepackage{bbold}
\usepackage{bbm}
\usepackage{url}
\usepackage[ruled, lined, linesnumbered, commentsnumbered, longend]{algorithm2e}
\usepackage{comment}
\newcommand{\RNum}[1]{\uppercase\expandafter{\romannumeral #1\relax}}

\newtheorem{proposition}{Proposition}

\newtheorem{assumption}{\textbf{Assumption}}
\newtheorem{lemma}{Lemma}

\newtheorem{definition}{Definition}

\usepackage{tabularx}
\usepackage{caption}
\usepackage{subcaption}

\title{Data-Driven Real-time Coupon Allocation in the Online Platform}

\author[1]{Jinglong Dai\footnote{The first three authors are co-first authors and contributed equally to this work.}}
\author[2]{Hanwei Li$^*$}
\author[3]{Weiming Zhu$^*$}
\author[1]{Jianfeng Lin}
\author[1]{Binqiang Huang}
\affil[1]{Meituan}
\affil[2]{City University of Hong Kong}
\affil[3]{The University of Hong Kong}

\date{}

\begin{document}
\maketitle

\small
\vspace{-1cm}
\begin{abstract}
\textbf{Problem definition:} Traditionally, firms have offered coupons to customer groups at predetermined discount rates. However, advancements in machine learning and the availability of abundant customer data now enable platforms to provide real-time customized coupons to individuals. In this study, we partner with Meituan, a leading shopping platform, to develop a real-time, end-to-end coupon allocation system that is fast and effective in stimulating demand while adhering to marketing budgets when faced with uncertain traffic from a diverse customer base. \textbf{Methodology/Results:} Leveraging comprehensive customer and product features, we estimate Conversion Rates (CVR) under various coupon values and employ isotonic regression to ensure the monotonicity of predicted CVRs with respect to coupon value. Using calibrated CVR predictions as input, we propose a Lagrangian Dual-based algorithm that efficiently determines optimal coupon values for each arriving customer within 50 milliseconds. We theoretically and numerically investigate the model performance under parameter misspecifications and apply a control loop to adapt to real-time updated information, thereby better adhering to the marketing budget. Finally, we demonstrate through large-scale field experiments and observational data that our proposed coupon allocation algorithm outperforms traditional approaches in terms of both higher conversion rates and increased revenue. As of May 2024, Meituan has implemented our framework to distribute coupons to over 100 million users across more than 110 major cities in China, resulting in an additional CNY 8 million in annual profit. \textbf{Managerial implications:} We demonstrate how to integrate a machine learning prediction model for estimating customer CVR, a Lagrangian Dual-based coupon value optimizer, and a control system to achieve real-time coupon delivery while dynamically adapting to random customer arrival patterns.

\end{abstract}

\section{Introduction}

Coupons play a crucial role in driving user engagement and revenue generation. Traditionally, firms such as Costco or McDonald's provide paper-based coupons to one or multiple groups of customers at pre-determined discount rates to stimulate demand. The effectiveness of coupons (\shortciteNP{sethuraman1992coupons}, \shortciteNP{spiekermann2011street}), together with the optimal allocation of marketing budget (\shortciteNP{holthausen1982advertising}, \shortciteNP{fischer2011dynamic}), have been extensively studied over the past decades. With recent advancements in machine learning, coupled with increased data availability and computing power, platforms such as Alibaba (\shortciteNP{zhao2019unified}), booking.com (\citeNP{albert2022commerce}), and Meituan started to harness the customer features and transaction records to personalize coupon values for each customer, aiming to engage their customers more effectively and allocate their marketing budgets more efficiently.

Companies used to adopt offline coupon allocation systems to prescribe personalized coupons. Such offline systems compute and store the optimal coupon value for all customers in a batched fashion and then push the value to the customers when they arrive. Take Meituan, a leading Chinese shopping platform, as an example. Meituan employed an offline Integer Program-based system (henceforth referred to as Offline-IP) to allocate coupons to each arriving customer. This system first utilizes off-the-shelf machine learning tools, such as XGBoost, to predict customer show-up probabilities and conversion rates based on historical user- and transaction-level data. The predicted outputs are then used by an offline integer program to determine the optimal coupon values for each customer in a batched fashion, while ensuring that total spending remains within the budget. Such an offline approach has three major disadvantages: (i) The company typically predetermines the coupon value for day $T$ at day $T-1$, and the decision-making process heavily relies on historical data and predictions rather than utilizing any information from day $T$. This could lead to prediction errors in the IP's coefficients and potentially cause misspecification of the optimization problem, leading to budget overruns and suboptimal allocation of coupon values. (ii) Jointly solving the IP for the entire customer base is computationally challenging. The system may need to group users based on the similarities of their conversion and show-up probabilities and then determine the optimal coupon value at the group level, making the prescribed coupon value suboptimal. (iii) The offline IP requires saving the coupon value assigned to each individual. Since large platforms typically have tens of millions of registered customers, storing all the coupon values can consume substantial memory.

In this work, we collaborate with Meituan Bike to design an end-to-end, \textit{online} coupon allocation system that aims to overcome the above-mentioned shortcomings of the offline allocation system. As the transportation division of Meituan, a leading shopping platform serving over 600 million users in China, Meituan Bike offers several payment methods for its users: pay-per-use or sign up for subscription plans. The company provides weekly and monthly subscription plans, each featuring distinct price points to cater to various user preferences. Similar to other platforms, Meituan Bike provides coupons to tens of millions of customers during the promotion period to incentivize them to sign up for one of its subscription plans while adhering to its marketing budget. Given that the monthly subscription plan is the most popular product which accounts for 90\% of sales in many cities, our work focuses on proposing a couponing system for the monthly subscription plan (i.e., a single-product setting). An example of Meituan's subscription plans and the coupon presentation is presented in Figure \ref{Fig:MT_Coupon} in Appendix \ref{app:snapshot}. 

Throughout the remainder of this paper, the terms ``online'' and ``real-time'' will be used interchangeably despite the subtle differences in their meanings. By ``online'', we refer to (1) the capability of determining a personalized coupon for customers upon their arrival using real-time information, as opposed to offering a pre-determined coupon value based on predicted customer show-up probabilities and (2) the capacity to dynamically adjust coupon values in response to the purchasing decisions of previous customers to better adhere to the marketing budget. When implementing a real-time allocation system, two significant challenges typically arise: strict time constraints for decision-making and budget management under fluctuating online traffic from a diverse customer base. To provide a seamless user experience, the system must determine and present the coupon value immediately upon a customer's arrival. For instance, Meituan's time limit for delivering coupon values to arriving customers is a mere 50 milliseconds. Such strict time requirements render conventional optimization methods, such as Mixed Integer Programming (MIP) or Dynamic Programming (DP), unsuitable in real time. As a result, platforms must strike a balance between the optimality of coupon values and rapid computation time.

Another critical challenge in the context of online platforms is budget management. Coupon allocation problems typically involve fixed budget resources, such as advertising costs in contract-based advertising (\shortciteNP{cheng2022adaptive},  \shortciteNP{chen2012ad}, \shortciteNP{wu2018budget}) or machine resources in computational resource allocation (\shortciteNP{jiang2020dcaf}). Traditionally, these problems are formulated as knapsack problems with fixed budget constraints \shortciteNP{hao2020dynamic} and solved offline. However, large online platforms need to dynamically monitor and adjust their spending levels while catering to millions of potential users who have uncertain show-up probabilities and purchase outcomes. As a result, traditional offline optimization models that rely heavily on predictions of user show-up probability often lead to budget overruns due to discrepancies between predicted and actual user arrivals.

In our work, we design an online, Lagrangian-dual method (LDM) based coupon allocation system capable of determining the optimal coupons for users almost instantaneously. Our coupon allocation system operates through a series of carefully structured steps. First, we input over 200 customer- and transaction-level features to train an XGBoost model, which estimates the customer conversion rates (CVR) under various coupon values. We address the non-monotonicity observed in the estimated CVR - a result of the complex data-generating processes within the platform - by applying isotonic regression (\shortciteNP{gupta2016monotonic}). Subsequently, the estimated CVR is input into a Lagrangian-dual-based optimization framework designed to determine the optimal coupon value in real time. Unlike the Offline-IP approach, which determines the optimal coupon value in a batched fashion (i.e., all the coupon values in day $T$ are jointly determined in $T-1$), LDM calculates the optimal coupon value for each individual upon their arrival. Meanwhile, as a crucial element in determining the optimal coupon value, the Lagrangian multiplier is first calculated through historical data and is then dynamically adjusted through a Proportional–Integral–Derivative (PID) controller (\shortciteNP{dong2009dynamic}) to account for real-time customer arrival pattern. This step ensures that the average value of distributed coupons adheres to the target level set by the marketing budget.

After developing our Lagrangian-dual-based coupon allocation system, we further study the performance of our proposed framework through observational data. We train and test our CVR model on 3 million user-level observational data entries. Our XGBoost-based prediction model, combined with Isotonic Regression, achieves an Area Under the Curve (AUC) of 89.13\% with a lower bias compared to the model without Isotonic Regression. Additionally, we demonstrate that our Proportional-Integral-Derivative (PID) control method keeps the average after-coupon price close to the predetermined budget with a marginal error below 1\%. Moreover, we compare the running time and the maximized objective value under LDM and Offline-IP by drawing on the user feature data and historical transaction records of more than 10 million users across four cities. We demonstrate that the Offline-IP runs out of memory for individual-level optimization, even for medium city size. We then segment customers into around 2,000 groups so both algorithms get to solve over a smaller problem set; our LDM performs 300 times faster. Finally, we find that the coupon values generated by the LDM are notably more dispersed than those by the Offline-IP. This is because LDM performs optimization at the individual level instead of the group level so that the model is able to more accurately pinpoint the optimal coupon value given the higher resolution of customer feature data.

Finally, we evaluate our proposed framework's effectiveness through large-scale online field experiments. Our field experiment encompasses over 3.4 million users from two major cities. Within each city, we deploy our Lagrangian-dual Method (LDM) to set the coupon values for the monthly subscription plan and compare our proposed allocation system against three other couponing systems previously used at Meituan, namely, Manual, Random, and Offline-IP approaches. We show through the experiment that when the objective is to maximize the total revenue, our proposed framework induces the highest conversion rate and results in the highest average revenue from the monthly subscription. Specifically, our proposed approach increases revenue from monthly subscriptions by 2.8\% per exposed user, translating into a CNY 6 million monthly revenue gain per 100 million exposed users. Moreover, we demonstrate that LDM is particularly effective in enhancing conversion rates for new, high-usage, and churned users.

Our work contributes to both existing literature and business practices in two ways. First, we design and introduce a real-time coupon allocation framework that can be readily used by practitioners. Specifically, we demonstrate that by combining PID feedback control with a dualized relaxed IP optimization, which is commonly used for static problems, we can effectively tackle dynamic budget allocation problems with minimal computation time. We offer a theoretical performance guarantee for the Lagrangian-Dual Method. Second, we conduct extensive field experiments in a high-stakes, real-world context and observe that implementing our framework can result in a substantial revenue uptick compared to traditional offline approaches. As of May 2024, Meituan Bike has implemented our framework to distribute coupons to over 100 million users in more than 110 major cities in China. Our proposed framework has increased revenue by 0.7\%, resulting in an additional 8 million CNY in annual profit for the platform compared to the traditional, offline IP approach, showing that our framework is effective in a competitive, real-time business environment. 

The remainder of this paper is structured as follows. Section 2 presents a review of the relevant literature. In Section 3, we provide an overview of the coupon allocation systems previously employed at Meituan alongside our newly proposed system. Sections 4 and 5 detail each step of our allocation system, describing our conversion rate prediction model and the optimization framework, respectively. We then assess our proposed framework's effectiveness through both offline and online experiments in Section 6. Finally, in Section 7, we conclude with a discussion of our findings and propose directions for future research.

\section{Literature Review}
Our work is related to the existing literature investigating coupons' impact on customer conversion and purchase decisions. Numerous prior studies have examined the effectiveness of coupons using observational data. For instance,  \citeANP{anderson2001sale} (\citeyearNP{anderson2001sale},\citeyearNP{anderson2001price}) demonstrate through mail-order catalog data that offering coupons can lead to mixed results in stimulating demand. Specifically, while coupons can potentially stimulate demand, their effectiveness in increasing demand diminishes when more items include them, and they may even jeopardize demand as they can be perceived as an indicator of inferior quality. However, more recent literature (\shortciteNP{spiekermann2011street}, \shortciteNP{reimers2019coupons}, \shortciteNP{smith2023optimal}) consistently provides evidence that coupons have a positive impact on demand. Nonetheless, the effect size can be moderated by other factors, such as firm characteristics and the proximity to the location where customers can redeem the coupons. The presence of these factors calls for a more personalized approach to maximize coupon effectiveness. Consequently, in our work, we propose a coupon allocation system that leverages customer attributes to generate tailored coupons for each individual customer. It is important to note that personalized coupons are distinct from dynamic pricing problems (e.g., \shortciteNP{chen2015recent} and \shortciteNP{den2015dynamic}) traditionally studied in operations literature, which primarily focus on industries such as airlines and ride-sharing. In these two industries, consumers are charged different prices based on factors such as the time of purchase or inventory availability. However, in our case, the platform issues varying coupon values primarily based on differences in customer features and transaction history.

Additionally, our paper studies the real-time coupon allocation problem with uncertain customer arrivals and fixed budgets. Thus, our work is related to the literature on budget allocation, which often appears in the context of advertisement (see \shortciteNP{chiu2018optimal} and \shortciteNP{holthausen1982advertising} for instance). In terms of techniques, as the system tries to determine not only if each customer should be included at all but also specific the coupon value prescribed to each customer, our coupon optimization problem is formulated as a variant of the Multiple-Choice Knapsack Problem, or MCKP (\shortciteNP{Kellerer2004}). Several previous studies (\shortciteNP{yan2023end}, \shortciteNP{albert2022commerce}) have adopted a similar setup to determine personalized coupon values. Our work, however, distinguishes from prior literature in the following ways: (1) we take the Lagrangian Dual of the MCKP problem; (2) we initialize the Lagrangian multiplier using historical data; and (3) we dynamically update the Lagrangian multiplier using a PID control approach. As a result, our system can prescribe optimal coupons one at a time for each arriving customer, effectively addressing the highly dynamic online market conditions, budget constraints, and stringent time limits.

Finally, we assess the effectiveness of our proposed coupon allocation algorithm through field experiments. In doing so, our work aligns closely with the extensive marketing literature investigating coupon distribution's causal effects via field experiments. In this context, \citeN{bawa1989analyzing} examines the impact of coupons on sales in a field experiment, comparing customers with and without purchase history for the advertised brand. \shortciteN{fong2015geo} conduct a field experiment to study how coupon usage is moderated by varying discount depths and the proximity of targeted individuals to the advertised movie theater. \shortciteN{sahni2017targeted} analyzes data from 70 field experiments, discovering that coupons can induce cross-category cross-selling and that their effect on sales can persist even after the coupon expires. \shortciteN{liu2021stimulating} identify behavioral factors, such as mental accounting and loss aversion, as crucial in designing coupons to stimulate economic recovery during the COVID-19 pandemic. Furthermore, several studies (\shortciteNP{heilman2002pleasant}, \citeNP{gopalakrishnan2021impact}) document the effectiveness of coupons in boosting revenue through field experiments, with underlying mechanisms ranging from encouraging unplanned purchases to attracting customers who make purchases without redeeming coupons. Our research diverges from prior literature in that we do not merely examine an established couponing scheme. Rather, we first develop a novel, LDM-based real-time coupon allocation system and then evaluate its performance against existing couponing systems through large-scale field experiments.


\section{Background and System Overview}
When customers request the bike-sharing service, they can choose between the following payment methods: (a) pay-per-use, or (b) sign up for subscription plans. The company offers weekly, monthly, and quarterly subscription plans with different price levels. Similar to other platforms, Meituan Bike provides coupons to tens of millions of customers during the promotion period to incentivize them to sign up for one of its subscription plans. Notably, the Monthly Subscription Plan is the most popular option, accounting for over 90\% of sales in many cities. Therefore, our paper focuses on the single-product setting and examines the coupon allocation problem for the Monthly Subscription Plan. As customers demonstrate significant heterogeneity and react differently to various coupon values, Meituan needs to customize the optimal coupon value for each arriving customer to maximize customer conversion rates while adhering to budget constraints. The following sections provide an overview of Meituan's previous Offline IP-based coupon allocation system. Next, we explain our newly developed system and highlight its advantages. Notably, the overall structure of both systems is quite general and can be adapted by other online platforms looking to implement their own real-time coupon systems.

\begin{figure}[H]  
\centering 
\includegraphics[width=0.9\textwidth]{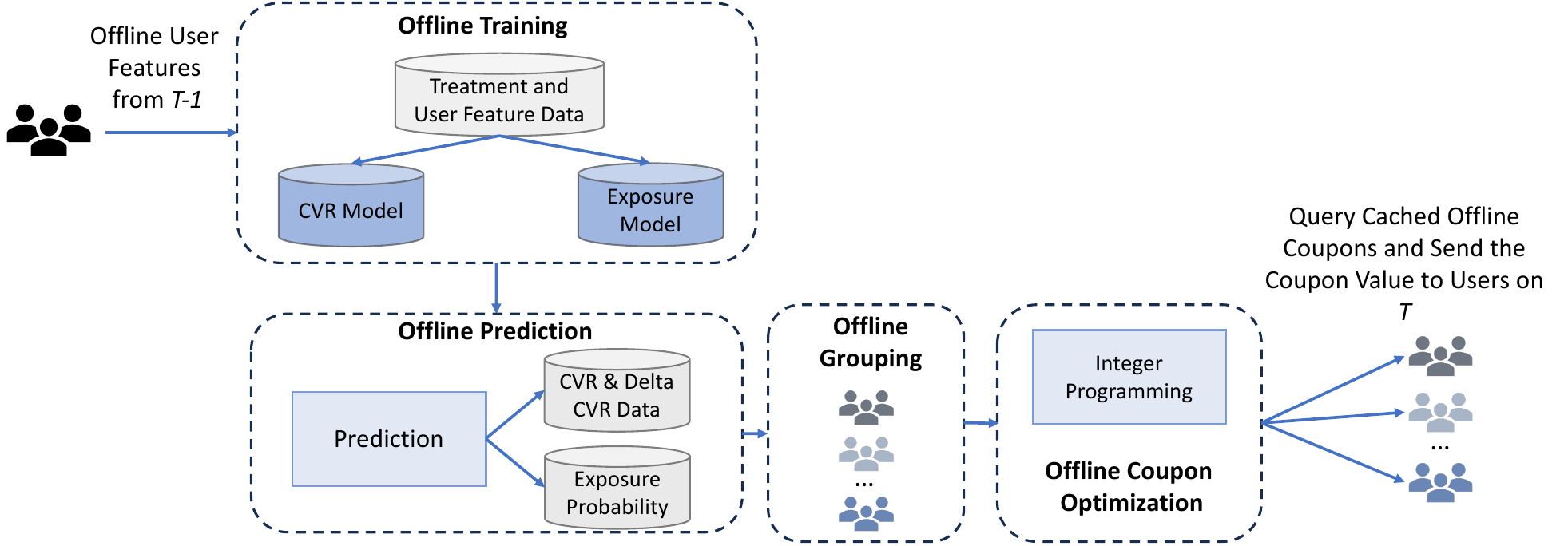} 
\caption{An overview of Meituan offline IP coupon allocation system.} 
\label{Fig:MT_Offline} 
\end{figure}

Meituan's coupon allocation systems can be roughly divided into two modules: a prediction module and an optimization module. Previously, Meituan adopted an offline, Integer Program (IP) based approach to the real-time allocation problem (denoted as Offline IP henceforth). As illustrated in Figure \ref{Fig:MT_Offline}, the coupon allocation system first uses off-the-shelf machine learning tools, such as XGBoost, to predict customer conversion rates (denoted as CVR henceforth) based on historical user-level and transaction-level data. Following this, the system estimates each customer's show-up probability and CVR under various coupon values drawn from a predetermined value ladder. The predicted CVR and show-up probability are then employed by an Integer Program (IP) to determine the optimal coupon value for each customer. However, due to computational burdens caused by the large customer population, the system has to first group customers based on the similarities of their CVR curves and show-up probabilities and then determine the optimal coupon value at the group level. Every individual within the same group would receive the same coupon amount (the formulation is presented in Section \ref{sec:optimization}). Finally, each customer's coupon value is stored and pushed to the customers in real time when they show up the following day.

\begin{figure}[H]  
\centering 
\includegraphics[width=0.8\textwidth]{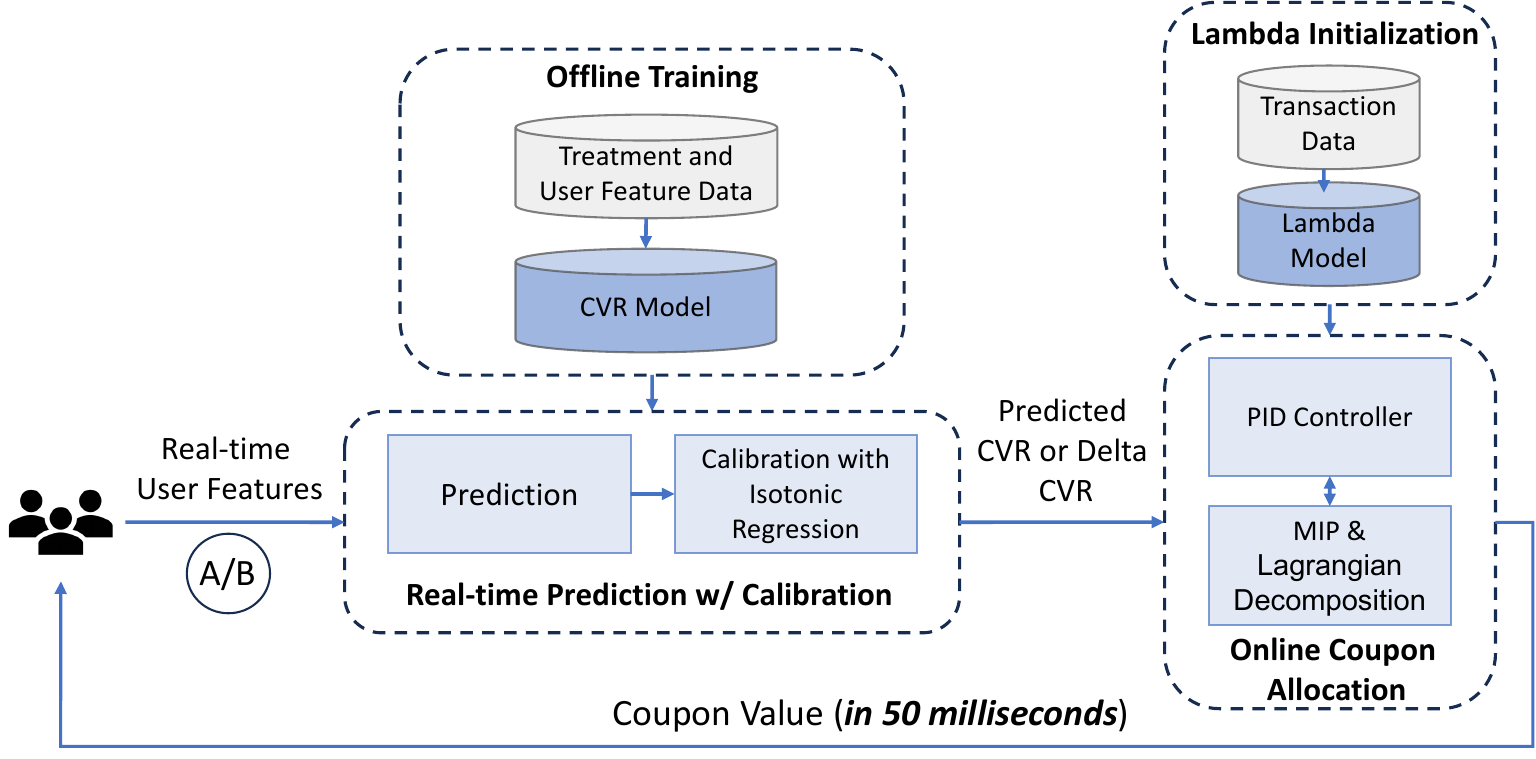} 
\caption{An overview of Meituan Offline-IP Coupon Allocation System.} 
\label{Fig:MT_RTB} 
\end{figure}

The Offline IP approach has three major flaws: (i) Due to computational limitations, the system needs to group users based on the similarities of their CVR and show-up probabilities and then determine the optimal coupon value at the group level, making the prescribed coupon value suboptimal. (ii) As Meituan predetermines the coupon value for day $T$ at day $T-1$, the decision-making process heavily relies on historical data and predictions rather than utilizing any real-time information from day $T$. This could lead to prediction errors in the IP problem's coefficients and potentially cause misspecification of the optimization problem. (iii) The offline IP requires saving the coupon value assigned to each individual. Since Meituan has more than 600 million registered customers, storing all the coupon values can consume substantial memory.

We address the aforementioned challenges by designing a fast, end-to-end framework that includes a series of offline and online algorithms capable of determining the optimal coupon value for users almost instantaneously upon their arrival. At its core, we propose a Lagrangian Dual-based algorithm that breaks down the population-level, batched optimization process into simple, streamlined problems for individual customers. Our framework can be summarized by Figure \ref{Fig:MT_RTB}, which includes the following steps:
\begin{enumerate}

    \item The Lagrangian multiplier $\lambda$ is initialized offline at the beginning of a given day $T$ by solving an offline relaxed Linear Program (IP) using an estimated arrival population(the formulation is also presented in Section \ref{sec:optimization}). The initialized $\lambda$ will be updated through the Proportional–integral–derivative (PID) controller during the day to reflect customer features in day $T$.

    \item Upon exposure, customer features are acquired and employed to forecast CVR under various coupon values using pre-established models. The estimated CVR is subsequently re-calibrated via Isotonic Regression to guarantee that the predicted CVR is non-increasing with respect to price.

    \item The predicted CVR and $\lambda$ are passed to the optimal coupon allocation module, which makes real-time decisions regarding the optimal coupon value for each user by solving a single equation (Equation~\eqref{eq:optimal_j}) generated by the Lagrangian Dual problem. During this process, the average after-coupon price for the targeted subscription plan must remain above a threshold, which is determined by the total marketing budget. Additionally, the Lagrangian multiplier $\lambda$ is updated in real time to ensure that the average discounted price remains higher than the predetermined lower bound.
    
    \item Finally, the coupon with the optimal amount decided online is pushed to the user and logged into the system.
\end{enumerate}

In the subsequent sections, we will discuss each step in more detail.

\section{Conversion Rate Prediction}
We start by introducing the training and calibration process for the customer conversion rate (CVR) model, which is the first step in the proposed coupon allocation system. Subsequently, we demonstrate the accuracy of the prediction algorithm using historical data.

\subsection{CVR Prediction}
The first step in determining the optimal coupon value for a customer is to predict the CVR under different coupon values. Assume there are a total of $J$ possible coupon values and $j \in \{1, 2, \dots, J\}$. We define $c_j$ and $p_j$ as the $j$th value in the pre-determined coupon value ladder and the respective price level. Specifically, $p_j=p_0-c_j$, where $p_0$ is the original price of the subscription plan. Note that it is possible that $c_j=0$, in which case the product is sold at its original price $p_0$ without any coupons. In the meantime, we denote $q_{ij}$ as the conversion probability of customer $i$ under price level $j$. In this way, we can train the conversion rate prediction model as 
\begin{equation*}
q_{ij}=f(\mathbf{X_i},p_j)+\epsilon_{ij}\,,
\end{equation*}
where $\epsilon_{ij}$ is the error term, $\mathbf{X_i}$ is a vector of customer features, including customer attributes, historical riding behaviors, and past coupon redemption records. 

As many features in $\mathbf{X_i}$ are derived through feature engineering, there can be a high correlation between variables (e.g., minimum, maximum, and mean bike daily usage time in the past 30 days). To handle such multicollinearity, $f$ is chosen from tree-based methods with L1 or L2 regularization, such as Extreme Gradient Boosting (XGBoost) or Light Gradient Boosting Machine (LightGBM). Once $f$ is trained on historical data, we can predict the conversion for a newly arriving customer $i$ at discount level $j$ as  $\tilde{q}_{ij} = f(\mathbf{X_i},p_j)$. The performance of the currently used predictive model is presented in Section \ref{sec:cvr_results}. Notably, CVR prediction is a modular component within the broader allocation system. This modularity allows for regular updates to the algorithm so that the state-of-the-art predictive methods can be integrated as needed.

\subsection{CVR Calibration with Isotonic Regression}
\label{sec.isotonic} 
When estimating customer conversion rates under different coupon values, we observe that it is not uncommon to obtain non-monotonic estimation results, in which case increased coupon value leads to decreased conversion rates. Multiple factors can contribute to such a non-monotonicity pattern. One such factor is the presence of unobserved confounding variables, which are primary sources of estimation biases in many contexts. However, as the prediction model we employ is trained using approximately 200 covariates, encompassing comprehensive customer characteristics as well as temporal and geo-spatial features (detailed in Section \ref{sec:experiments}), the impact of unobserved confounding variables should be minimal in our setting. Another potential cause of the non-monotonicity pattern is the non-parametric nature of machine learning models (e.g., XGBoost), which can be exacerbated when customers are swayed by simultaneous subsidy campaigns orchestrated by various departments within the company. To demonstrate the magnitude of this impact, we construct a simple numerical simulation model below.

\begin{table}[t]
\small
\centering
\caption{Simulation on Violations of Monotonicity in CVR}
\begin{tabular}{lcc}
\hline
\textbf{Scenarios} & \textbf{Covered Population} & \textbf{\% of Non-monotonic CVR} \\
\hline
Basic      &  0\% &  2.34\%     \\
Campaign-Low  & 10.35\%    &  6.84\%      \\
Campaign-Medium &  25.03\%    &  10.95\%      \\
Campaign-High   &  71.90\%  &   69.95\%     \\
\hline
\end{tabular}\label{tab:cvr_sim}
\end{table}

Our simulation contains 100,000 heterogeneous customers. Customer features $\mathbf{X_i}$ include base utility and price elasticity, both of which are drawn from two independent distributions (the detailed simulation setup is presented in Appendix \ref{app:cvr}). We consider a simple setting where there are no unobserved noises ($\epsilon_{ij} = 0$), and all customer features $\mathbf{X_i}$ are perfectly known by the company. The provided prices, or equivalently, the coupon values, are generated randomly from a predetermined price ladder. Customers proceed with a purchase when their utility exceeds a certain threshold. 

Upon observing customers' purchase decisions, we employ an XGBoost model to predict conversion ($\tilde{q}_{ij}$) for each customer using $\mathbf{X_i}$ and prices (coupon values). This allows us to recover the relationship between $\tilde{q}_{ij}$ and coupon values and determine if the conversion is monotonically decreasing (increasing) with respect to offered prices (coupon values). Additionally, we simulate scenarios in which prices from the concurrent campaign are not randomly generated but specifically designed to offer lower prices to customers with low willingness to pay in order to incentivize purchases. In each scenario, a varying percentage of the customer population is targeted.

The simulation results in Table \ref{tab:cvr_sim} illustrate the average frequency of non-monotonic CVR patterns over 30 iterations of simulations. Firstly, we discover that 2.34\% of customers exhibit non-monotonic demand curves even in the basic setting where all prices are generated randomly, which contradicts the underlying data-generating process. We also observe that a concurrent marketing campaign targeting a specific customer segment can significantly increase instances of non-monotonic CVR predictions. Moreover, these cases become more common as the campaign coverage broadens.

At Meituan, it is common for different business units to run concurrent campaigns targeting specific customer groups. Consequently, the CVR estimated using historical or experimental data can exhibit nonmonotonic patterns with respect to price (or coupon value). These nonmonotonic patterns are unlikely to be consistent with the underlying data-generating process, and, once used as input for the optimization problem, can bias the prescribed coupon value and ultimately lead to a suboptimal conversion rate. Thus, it is essential to recover monotonicity in the predicted CVR ($\tilde{q}_{ij}$) to avoid biased outcomes. Notably, many machine learning models, such as XGBoost and certain neural networks, have the option to force monotonicity constraints. However, platforms like Meituan may frequently experiment with and implement various machine learning models for predictions, and some of these methods may not offer the option to impose monotonicity constraints. To address this, we incorporate a post-processing module after the CVR prediction step. Specifically, we apply Isotonic Regression (IR) \cite{barlow1972isotonic} to the predicted CVR to correct any non-monotonic distortions using the following equation:
\begin{equation*}
\begin{aligned}
\min \ &\sum_{q_{ij}, \forall j\in J} w_i (\hat{q}_{ij} -\tilde{q}_{ij'})^2 \\
s.t. \ & \hat{q}_{ij} \ge \hat{q}_{ij'} \quad \forall (j, j') \in P 
\end{aligned}
\end{equation*}
where $\tilde{q}_{ij}$ and $\hat{q}_{ij}$ are IR-uncorrected and IR-corrected estimated conversion rate for customer $i$ under coupon value $j$. $P={(j,j'):\hat{q}_{ij} \ge \hat{q}_{ij'}}$ specifies the partial ordering of the coupon level inputs set.


\begin{figure}[t]  
\centering 
\includegraphics[width=1.0\textwidth]{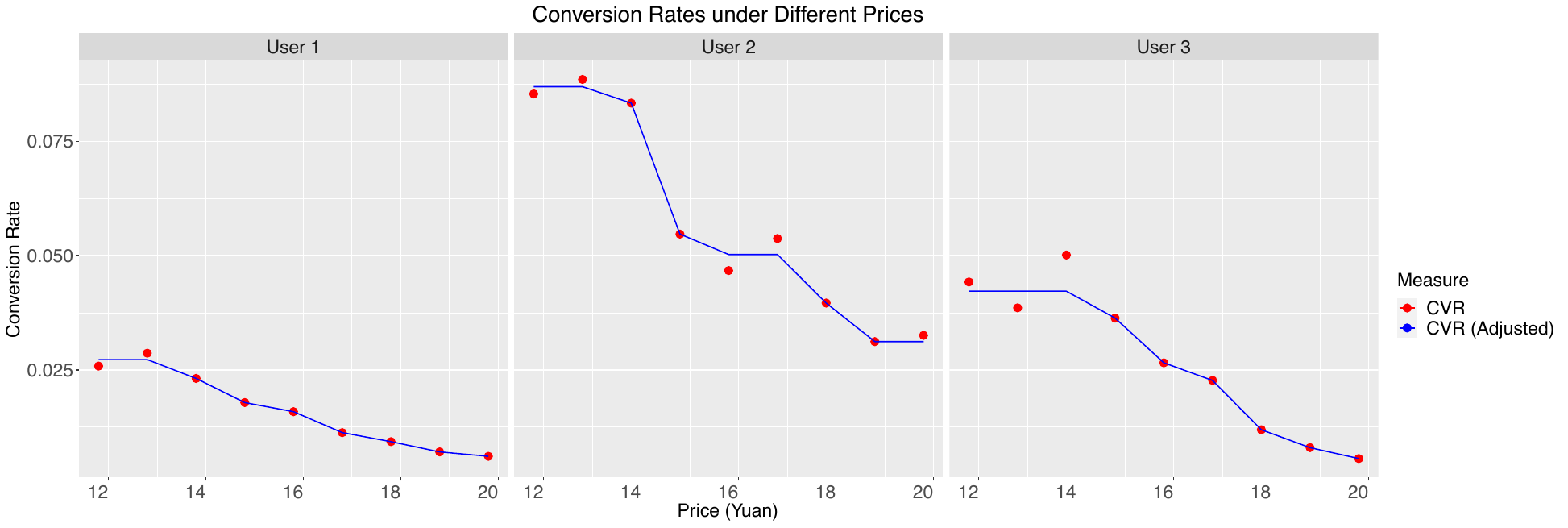} 
\caption{Conversion Rates and IR-Adjust Conversion Rates under Different Prices.} 
\label{Fig:IR} 
\end{figure}

As the IR step is independent of the base estimation model, the company can easily adjust the calibration step without affecting the training function $f$. To understand how the IR step smooths out local positive elasticity in CVR estimation, we plot the estimated CVR and the IR-adjusted CVR in Figure \ref{Fig:IR}. Notably, User 2 in Figure \ref{Fig:IR} implies that, when $(j, j')\in P $ but $\tilde{q}_{ij} \le \tilde{q}_{ij'}$, isotonic regression does not simply equate $\hat{q}_{ij'}$ to $\tilde{q}_{ij}$. Instead, it computes the value of $\hat{q}_{ij}$ and $\hat{q}_{ij'}$ that minimizes the squared difference between the corrected and uncorrected CVR. 

\subsection{Evaluation of the CVR Prediction Model}\label{sec:cvr_results}
In this section, we evaluate the performance of the CVR prediction model on historical data. The model is trained using three million historical user data entries. The dataset is sourced from the random traffic group, which accounts for 20\% of the traffic from around 100 cities in China. In the random traffic group, all coupon values in the value ladder are distributed to users with equal probabilities. This large data size and random traffic distribution significantly improve identifiability and reduce the risk of overfitting. We allocate 80\% of the data for model training and 20\% for testing to assess model performance. The models are trained using over 200 features, encompassing user attributes, riding behavior, and historical transaction records.

Using customer features and transaction data from the random traffic dataset, we train an XGBoost model to predict CVR. Specifically, using two metrics, we compare the out-of-sample performance between the traditional XGBoost model and the XGBoost model with isotonic regression (XGB-IR). The first metric is the Area under the ROC Curve (AUC), which measures the model's ability to predict customer purchases with no-purchase decisions. An AUC value of 0.5 implies that the model performs no better than random guessing, while an AUC value of 1.0 indicates that the model perfectly distinguishes between the purchase and no purchase results. In our case, the AUC metric for both models is 0.891, which shows good prediction power.

In addition to AUC, we also use Predict Conversion Over Conversion (PCOC)(\shortciteNP{wu2024metasplit}). PCOC is defined as the ratio of the CVR to the realized binary outcome of whether the customer makes a purchase, which can be formally expressed as:
\begin{equation}
    PCOC = \frac{\sum_i \hat{q}_{ij^*}}{\sum_i \mathbb{1}_{ij^*}}
\end{equation}
Intuitively, PCOC measures the model's overall prediction bias. A value greater than 1 indicates an overestimation of customer conversion, while a value less than 1 suggests an underestimation. After applying isotonic regression to the XGBoost output, we find that the PCOC improves from 0.971 to 0.982. The results indicate that the IR-adjusted CVR not only aligns better with business intuition but also reduces the model's prediction bias. Consequently, it serves as a better input to the subsequent optimization module compared to the unadjusted CVR.

\section{Optimization Formulation}\label{sec:optimization}
The previous section describes the estimation and calibration procedures used for CVR predictions. The predicted conversion rates serve as inputs for an optimization framework, which determines the optimal coupon value for each incoming customer in real time while adhering to budget constraints. In line with the conversion rate notation, we use subscript $i$ to represent the customer index and subscript $j$ to represent the coupon level. We denote $v_{ij}$ as a general term to be maximized for customer $i$ under coupon level $j$. Depending on the company's specific goals, $v_{ij}$ can be either revenue, conversion rate, or other metrics as the objective. We define $I_0$ as the entire customer population, that is, all registered app users for the city. Notably, while the size of $I_0$ can be extremely large (over millions for large cities), only a subset $I \subseteq I_0$ will use the app service each day. We refer to $I$ as the exposed or show-up population to whom real-time coupon offers must be made. Lastly, we use $|I|$ and $|I_0|$ to indicate the cardinality (the number of customers) of $I$ and $I_0$, respectively. We introduce the previous and proposed coupon frameworks in the following subsections.

\subsection{Offline Integer Programming Formulation}\label{sec:offline_ip}

In this section, we demonstrate the Offline Integer Programming (IP) formulation previously used by the company. We begin by discussing the problem's budget constraint.

When a company is to determine the value of coupons, the first step typically involves establishing a budget constraint to avoid excessive discounts. To achieve this, traditional offline retailers often employ a deterministic, pre-determined budget cap. Let $p_{j}$ represent the discounted price of the subscription plan given coupon level $j$, and $\mathbb{1}_{ij}$ is the binary realization of whether coupon level $j$ is offered and ultimately purchased by customer $i$. The traditional type of budget constraint can then be expressed as $\sum_{i\in I}\sum_{j\in J} (p_0 - p_{j})  \mathbb{1}_{ij} \leq B$, where $p_0$ refers to the undiscounted price of the subscription plan and $B$ denotes the total marketing budget. Since the purchase realization $\mathbb{1}_{ij}$ cannot be known in advance, we further replace $\mathbb{1}_{ij}$ by $\hat{q}_{ij}x_{ij}$, where $\hat{q}_{ij}$ is the estimated conversion rate and $x_{ij}$ the binary decision variable indicating if coupon level $j$ is offered to customer $i$. This way, the traditional budget constraint is expressed as $\sum_{i\in I}\sum_{j\in J} (p_0 - p_{j}) \hat{q}_{ij} x_{ij} \leq B$.

For online platforms, adhering to a predetermined marketing budget can be challenging due to the volatility and unpredictability of customer arrivals. When a platform determines coupon values based on a specific set of customers $I$, additional customers outside of $I$ (e.g., tourists whose IP addresses are not within the city) may not receive any coupons to avoid violating the budget constraint, which results in a suboptimal coupon allocation. To address this volatility in customer arrival, the company adopted an alternative definition for the budget constraint that requires the average price level (or the average coupon discount) to be higher (lower) than a certain threshold. Specifically, the budget constraint can be expressed as follows:
\begin{equation}\label{eq:mip_budget}
    \frac{\sum_{i\in I}\sum_{j\in J} p_{j}\; \hat{q}_{ij}\; x_{ij} }{\sum_{i\in I}\sum_{j\in J} \hat{q}_{ij}\;x_{ij} } \geq p_b,
\end{equation}
Equation \eqref{eq:mip_budget} requires that the expected average offered price cannot fall below $p_b$ -  a lower bound on the price that is exogenously set by the company's marketing team based on a range of factors such as the marketing budget and operating cost. However, as the arriving population $I \subseteq I_0$ is unknown beforehand, we require the average offered price to stay above $p_b$ in the expectation form instead. To achieve this, the company needs to estimate the individual show-up probability $\hat{s}_{i}$ for all potential customers $i \in I_0$ The budget constraint is then expressed as the following: 
\begin{equation}\label{eq:arpu_constraint}
\frac{\sum_{i\in I_0}\sum_{  j\in J} p_{j} \;\hat{q}_{ij}\;\hat{s}_{i}\; x_{ij} }{\sum_{i\in I_0}\sum_{ j\in J} \hat{q}_{ij}\;\hat{s}_{i}\; x_{ij}} \geq p_b
\end{equation}
With the budget constraint defined, we describe an offline Integer Program (IP) for coupon value determination. The existence of the budget constraint Equation \eqref{eq:arpu_constraint} ties all potential customers together. Therefore, solving the IP in real time (under 50 milliseconds) for each individual arrival is impossible. Since any potential customer $i \in I_0$ may show up, the company needs to solve and store for the offline IP involving all potential customers $i \in I_0$ one day before. Specifically, the offline IP can be written as follows:  
\begin{equation}
\label{eq:offline_ip}
\begin{aligned}
\max_{x_{ij}} \ &\sum_{i\in I_0}\sum_{  j\in J}	\hat{s}_{i} \; v_{ij} \; x_{ij} \\
s.t. \   & 
\frac{\sum_{i\in I_0}\sum_{j\in J} p_{j} \;\hat{q}_{ij}\;\hat{s}_{i}\; x_{ij} }{\sum_{i\in I_0}\sum_{ j\in J} \hat{q}_{ij}\;\hat{s}_{i}\; x_{ij}} \geq p_b\\
 & \sum_{j\in J} x_{ij} = 1 \quad \forall i \in I_0 \\
&  x_{ij} \in \{0,1\}\quad\forall i \in I_0, j \in J\\
\end{aligned}
\end{equation}
In the objective function, $v_{ij}$ is a general term that is shaped by the value generated from customer $i$ purchasing coupon level $j$. We keep $v_{ij}$ general to reflect that companies at different stages may have different targets, ranging from maximizing total conversion rate to total revenue. Later in Section \ref{sec:experiments}, we assume $v_{ij}$ represents individual revenue, which is defined as $v_{ij}=p_j \hat{q}_{ij}$, to stay consistent with the company's actual objective. 

Practically, the offline IP approach can be inconvenient in multiple ways. First, given the MIP formulation, the company can only determine the coupon values for the entire customer base (i.e., in a batched fashion) rather than individually calculating the optimal coupon value for each customer in real-time as they arrive (i.e., in a streamlined fashion). This is why Meituan had to pre-compute the optimal coupon value on day $T-1$ and use it on day $T$. This introduces the need to forecast the show-up probability $\hat{s}_i$ for each customer, which in turn brings prediction error and misspecification into the IP formulation. Second, following the previous point, since the coupons are pre-determined and stored, the system cannot utilize the real-time customer information when they actually arrive or when their profile is updated. Third, large platforms such as Meituan serve hundreds of millions of customers, directly solving Formulation \eqref{eq:offline_ip} can cause computers to run out of memory. As a result, Meituan had to group customers into clusters based on their feature similarities and determine coupon values at the group level. This means that to reduce the problem size in Formulation \eqref{eq:offline_ip}, the company has to model customers on a group level so that index $i$ represents customer groups instead of individuals. The process of grouping reduces data resolution and can result in suboptimal coupon values that potentially violate the budget constraint and diminish customer conversion rates.

\subsection{Lagrangian Dual Method}\label{sec:LDM}
In the Offline-IP approach, the optimization has to be carried out over the entire population $I_0$ to ensure the platform stays within its budget limit. The resulting coupon values from Offline-IP can be significantly suboptimal due to the estimation errors stemming from not only $\hat{q}_{ij}$ but the customer's show-up probability $\hat{s}_{ij}$, as well as the efficiency loss in the customer grouping process.

To address the aforementioned challenges, we propose to apply the Lagrangian Dual Method (LDM). In the LDM approach, we approximate the arrival population on day $T$, denoted as $I$, using $\hat{I}$, which is constructed based on the arrival patterns from days $T-1$ to $T-T_0$. The value of $T_0$ is determined by the company through trial-and-error. Since the size and nature of customers may differ between day $T$ and previous days, real-time transaction information will be used to update the prescribed coupon values to adhere to the budget constraint. As we will demonstrate later in this subsection, our LDM approach eliminates the need to estimate $\hat{s}_{ij}$ and, more importantly, enables the determination of optimal coupon allocation at the individual level upon each customer's arrival. We formulate the offline coupon optimization as follows:
\begin{equation}
  \label{eq:basic_formulation}
\begin{aligned}
\max_{x_{ij}} \ &\sum_{i\in \hat{I}}\sum_{  j\in J}	v_{ij} \; x_{ij} \\
s.t. \   & 
 \frac{\sum_{i\in \hat{I}}\sum_{j\in J} p_{j} \;\hat{q}_{ij}\; x_{ij} }{\sum_{i\in \hat{I}}\sum_{ j\in J} \hat{q}_{ij}\; x_{ij}} \geq p_b\\
 & \sum_{j\in J} x_{ij} = 1 \quad \forall i \in \hat{I} \\
&  x_{ij} \in \{0,1\}, \forall i \in \hat{I}, j \in J\\
\end{aligned}
\end{equation}

Comparing the proposed new offline Formulation \eqref{eq:basic_formulation} with the traditional offline Formulation \eqref{eq:offline_ip}, we eliminate the show-up probabilities $s_i$ and update the entire population $I_0$ to the estimated show-up population $\hat{I}$. As a consequence, we can no longer pinpoint the optimal coupon values for all potential customers $i \in I_0$. To address this, we will relax Formulation \eqref{eq:basic_formulation} into the Linear Programming (LP) form (the relaxation gap of which will be discussed in the next subsection) and then dualize the budget constraint for the Lagrangian decomposition. We denote $\lambda$ as the multiplier associated with the budget constraint and $\mu_i$ as the multiplier associated with the coupon selection constraint $i$. To facilitate analysis, we also invert the sign of the objective function and reformulate \eqref{eq:basic_formulation} in the following Lagrangian min-max form.
\begin{equation}
\begin{aligned}
\label{eq:lag_lp}
\min_{x_{ij}}\max_{\lambda\geq 0, \mu_i} L(x_{ij}, \lambda, \mu_i) 
&= \sum_{i\in \hat{I}}\sum_{ j\in J}- v_{ij}  x_{ij}  + \lambda (\sum_{i\in \hat{I}}\sum_{j\in J}   \hat{q}_{ij} x_{ij} p_b - \sum_{i\in \hat{I}}\sum_{ j\in J} \hat{q}_{ij} x_{ij}p_{j}   )
+\sum_{i\in \hat{I}}(\mu_i (\sum_{j\in J} x_{ij} - 1 )) \\
&= \sum_{i\in \hat{I}}\sum_{ j\in J}x_{ij} (-v_{ij}+ \lambda \hat{q}_{ij} (p_b - p_{j}))
\end{aligned}
\end{equation}
Using the strong duality of LP, we establish that $L$ must be equal to the objective of Formulation \eqref{eq:basic_formulation}. Furthermore, since the $\min_x$ and $\max_{\lambda,\mu}$ operations are interchangeable, and the terms associated with the multiplier $\mu$ negate each other, we arrive at the following formulation.
\begin{equation}
\label{eq:max_min_L}
\max_{\lambda\geq 0} \min_{x_{ij}}L(x_{ij}, \lambda)=\min_{x_{ij}}\max_{\lambda\geq 0} L(x_{ij}, \lambda)= \sum_{i\in \hat{I}}\sum_{ j\in J}x_{ij} (-v_{ij}+ \lambda \hat{q}_{ij} (p_b - p_{j}))
\end{equation}
The problem then can be decomposed into $|\hat{I}|$ individual-level optimization problems, where we need to decide on $x_{ij}$ given a fixed value of $\lambda$. As there are only a few discrete levels of coupon levels $j \in J$, for each arriving customer $i$, we can simply plug in each coupon value and determine the optimal coupon level $j^*$ given $\lambda$ and the estimated conversion rate $\hat{q}_{ij}$ using the following equation:
\begin{equation}\label{eq:optimal_j}
j^*(\lambda) = \mathrm{argmax}_{j\in J}{(v_{ij} - \lambda \hat{q}_{ij} (p_b - p_{j}))}
\end{equation}
When there are multiple optimal solutions to Equation \eqref{eq:optimal_j}, we will take the solution with the lowest coupon value (or highest price) to ensure the budget constraint in Formulation \eqref{eq:basic_formulation} is satisfied. According to Equation \eqref{eq:optimal_j}, given a fixed value of $\lambda$, the determination of the optimal coupon value boils down to comparing predetermined coupon values for each customer. This eliminates the need to solve the optimization problem for the entire customer base every time a new customer arrives. At the same time, we can utilize the real-time customer information to update input parameters $v_{ij}$ and $\hat{q}_{ij}$ to make the coupon offering process more accurate and, therefore, more likely to be optimal. With the LDM approach, the determination of the optimal coupon value can be completed within 50 milliseconds, enabling the company to push coupon values to the arriving customers in real time. 

After knowing how to solve for $x_{ij}$, we need to find the optimal Lagrangian multiplier. Notably, as we lack information on $I$ and use estimated population $\hat{I}$ in Formulation \eqref{eq:basic_formulation}, we denote the optimal offline Lagrangian multiplier as $\hat{\lambda}$. As the objective $L$ is piecewise linear in $\lambda$ and takes the max-min form in Equation \eqref{eq:max_min_L}, it is concave in $\lambda$. Therefore, we can determine the value of $\lambda$ through trisection or golden section search. We present a simple trisection algorithm below in Algorithm 1 to determine the optimal $\hat{\lambda}$ for the offline Formulation \eqref{eq:basic_formulation}.

\begin{algorithm}[h!]\small
\caption{Offline Trisection Search for Optimal $\hat{\lambda}$}\label{algo_trisection}
\SetAlgoLined
\SetKwInOut{Input}{input}\SetKwInOut{Output}{output}
\SetKwInOut{Init}{init.}
\Input{ lower bound $\lambda_{low}$, upper bound $\lambda_{high}$, tolerance level $\epsilon$, CVR $\hat{q}_{ij}$, value function $v_{ij}$}
\Init{$\lambda_1$ $= \lambda_{low}$, $\lambda_4 = \lambda_{high}$, $L_1 = L_4 = -\infty$,}
\While{$\lambda_4-\lambda_1 >\epsilon$}
{Compute the trisection points $\lambda_2 = \lambda_1 + \frac{1}{3}(\lambda_4-\lambda_1 )$, $\lambda_3 = \lambda_1 + \frac{2}{3}(\lambda_4-\lambda_1 )$\\
Set $L_2 = L_3 = 0$\\
\For{$i\in \hat{I}$}{
Find $j^* = \mathrm{argmax}_{j\in J}{(v_{ij} - \lambda \hat{q}_{ij} (p_0 - p_{j} ))}$\\
Update $L_2 = L_2 + \sum_{i\in I} (v_{i{j^*}}- \lambda_2 \hat{q}_{i{j^*}} (p_b - p_{i{j^*}}))$\\
Update $L_3 = L_3 + \sum_{i\in I} (v_{i{j^*}}- \lambda_3 \hat{q}_{i{j^*}} (p_b - p_{i{j^*}}))$\\
}
\eIf{$L_2\geq L_3$}
{Set $\lambda_4=\lambda_3$, $L_4 = L_3$\\}
{Set $\lambda_1=\lambda_2$, $L_1 = L_2$\\}
}
\Output{$\hat{\lambda} = \frac{\lambda_1+\lambda_4}{2}$}
\end{algorithm}

To this end, we do not need to plug the value back into Equation \eqref{eq:optimal_j} to recover the optimal coupon level $j^*$, because the solution for the offline estimated population $\hat{I}$ will not be useful for the real-time population $I$. Nevertheless, the multiplier $\hat{\lambda}$ is crucial as leverage in the real-time coupon allocation process. The company uses offline inputs and estimated population $\hat{I}$ to solve for \eqref{eq:basic_formulation} one day before. The obtained $\hat{\lambda}$ will be used for Equation \eqref{eq:optimal_j} to determine the optimal coupon level in real time. This approach does not require solving any optimization problem in real time and, therefore, satisfies the speed requirement of 50 milliseconds for each arriving customer. Still, it only works well when $\hat{\lambda}$ is a good approximation to the ground truth $\lambda$. Next, we first provide some theoretical properties of the LDM approach in subsection \ref{sec:theory}. Then, we introduce the real-time monitoring mechanism to guide $\hat{\lambda}$ in subsection \ref{sec:pid}.

\subsection{Theoretical Properties of the LDM-Based Approach}
\label{sec:theory}
Notably, in determining the optimal coupon value $j^*$, we leverage the similarity between $\hat{I}$ and $I$ to compute the Lagrangian multiplier $\hat{\lambda}$ and determine the corresponding $j^*(\hat{\lambda})$. This approach enables us to determine coupon values even for customers who do not belong to the estimated population, i.e., $i \notin \hat{I}$. However, it is crucial to ensure that $\hat{\lambda}$ is close to $\lambda^*$ so that the total customer conversion rates are maximized and the budget constraint is satisfied. To this end, we first theoretically demonstrate that the aforementioned statement holds asymptotically when the constructed population $\hat{I}$ is close to the actual population $I$. Later, in Section \ref{sec:pid}, we propose solutions for scenarios where the assumption does not hold, and $\hat{\lambda}$ deviates from $\lambda^*$ due to nonstationary customer demographics across days. We start the asymptotic discussion by making the following assumption.

\begin{assumption}
\label{assump:arrival}
Each potential customer $i \in I_0$ has a non-zero show-up probability $s_i>0$ every day.
\end{assumption}
Assumption \ref{assump:arrival} holds when, during the period of interest, customers do not churn and maintain a relatively stable show-up probability across days. Since the show-up probabilities $s_i$ can differ between weekdays and weekends, or vary due to factors such as temperature and weather, the company needs to carefully select previous days with similar features to approximate $I$. Another implication of Assumption \ref{assump:arrival} is that the subscription sign-up and expiration rates should remain similar so that a customer who purchases the subscription plan is replaced by an identical customer whose subscription plan has just ended. 
\begin{proposition}
\label{prop_I_gap}
Under Assumption \ref{assump:arrival}, as the total population size approaches infinity ($|I_0| \to \infty$), the Lagrangian multiplier computed using the average of previous days' $\hat{I}$ converges to the ground truth, i.e., $\hat{\lambda} \to \lambda^*$. Furthermore, the optimal coupon levels satisfy $j^*(\hat{\lambda})=j^*({\lambda}^*)$.
\end{proposition}
Proposition \ref{prop_I_gap} suggests that the gap between the $\hat{\lambda}$ and $\lambda^*$ will diminish given a sufficiently large population and carefully chosen $\hat{I}$. This implies that such approximation works especially well for large cities with millions of users. For these markets, the estimated $\hat{\lambda}$ can serve as a reasonably good approximation for the next day's coupon allocation process. In Section \ref{sec:pid}, we discuss how $\hat{\lambda}$ can be updated throughout the day to account for the approximation error between $\hat{I}$ and $I$. Additionally, we conduct robustness checks in the finite sample case to investigate the impact of any misspecification of $\hat{\lambda}$ on the optimal coupon offered.

Next, we investigate the impact of $\hat{\lambda}$ on the resulting price and feasibility of the solution. To this end, we denote the coupon optimization problem formulated with the ground truth customer arrival $I$ as $\text{IP}_0$. The formulation of $\text{IP}_0$ is identical to that of Formulation \eqref{eq:basic_formulation}, except that the budget constraint is formulated using real time arriving population $I$, as $\sum_{i\in I}\sum_{j\in J} p_{j} \;\hat{q}_{ij}\; x_{ij}/\sum_{i\in I}\sum_{ j\in J} \hat{q}_{ij}\; x_{ij} \geq p_b$. If we denote the LP relaxation of $\text{IP}_0$ as $\text{LP}_0$, it is clear that $\lambda^*$ is the optimal Lagrangian multiplier for the $\text{LP}_0$. Also, denote $p_{j^*(\hat{\lambda})}$ the offered price under coupon level $j^*(\hat{\lambda})$, and we will have the following directional results.
\begin{proposition}
\label{prop:lambda}
Given $\lambda^*, \hat{\lambda} > 0$, the offered price $p_{j^*(\hat{\lambda})}$ is non-decreasing in $\hat{\lambda}$ for all $i \in \hat{I}$. In addition, if $\hat{\lambda} < \lambda^*$, $\text{LP}_0$ will be infeasible. If $\lambda > \lambda^*$, the budget constraint of $\text{LP}_0$ will not bind.
\end{proposition}
Proposition~\ref{prop:lambda} sheds light on the impact of $\hat{\lambda}$ on the offered price and marketing spending. In particular, when $\hat{\lambda}$ is overestimated, Equation \eqref{eq:optimal_j} tends to recommend a higher coupon value (leading to a reduced offered price), which may cause the company to surpass its marketing budget. On the other hand, if $\hat{\lambda}$ is underestimated, the company will prescribe a more conservative coupon value, resulting in reduced marketing spending and a lower overall conversion rate.

Finally, as $j^*(\hat\lambda)$ is generated from the LP relaxation of the original IP problem \eqref{eq:basic_formulation}, it is thus important to understand the gap between the optimal objective values of the original problem and its LP relaxation. The asymptotic LP relaxation gap is presented through Proposition \ref{prop:lp_gap} under the following assumption.

\begin{assumption}
\label{assump:feature}
Customers' input CVR $\hat{q}_{ij}$ and value function $v_{ij}$ are randomly drawn from continuous distributions.
\end{assumption}
\begin{proposition}
\label{prop:lp_gap}
Under assumption \ref{assump:feature}, an optimal solution with at most one pair of fractional solutions exists for the relaxed LP from Formulation \eqref{eq:basic_formulation}. 
\end{proposition}

Assumption \ref{assump:feature} holds when the dimensionality of the customer feature space is sufficiently large, and the estimation model maps these features onto a continuous outcome space. Typically, Meituan employs approximately 200 distinct features, including user attributes and ride behavior, to inform its predictions. This granularity reduces the likelihood that two customers will possess identical sets of CVR and value function predictions. Consequently, it is statistically unlikely for multiple customers to exhibit the same set of fractional solutions simultaneously, thus ensuring a tight LP relaxation when the population size is large. However, in practice, the prediction model returns customer CVR with finite decimals, which may allow for the presence of more than one customer with fractional solutions.  To investigate the presence and magnitude of this issue, we conduct robustness checks in Section \ref{sec:pid}, using the existing $\hat{q}$ and $v$ forecast models and the LDM to determine coupon values. Surprisingly, these checks reveal no fractional solutions in the LP relaxation across our numerical experiments and provide numerical evidence of the tightness of the LP relaxation.

\subsection{Online Control of Multiplier $\lambda$}\label{sec:pid}
As mentioned in the previous subsection, the Lagrangian multiplier $\lambda$ serves as a lever to control the average discounted price. Ideally, the estimated Lagrangian multiplier $\hat{\lambda}$ should be close to the ground truth $\lambda^*$, as Proposition \ref{prop_I_gap} suggests. However, in practice, Assumption \ref{assump:arrival} may not always hold since the show-up population can be nonstationary across days. For example, customers on weekends and holidays may exhibit very different purchasing patterns, causing the estimated $\hat{\lambda}$ to deviate from $\lambda^*$. Furthermore, the estimated conversion rate $\hat{q}_{ij}$ may not accurately predict customer purchasing behavior. When combined with an imprecise $\hat{\lambda}$, this can introduce significant misspecification into Equation \eqref{eq:optimal_j}. Consequently, if we denote the average offered price for realized demand up to time $t$ as $p_t$, where $p_t = \sum_{i\in I_t} p_{j^*} \; \mathbb{1}_{ij^*}/\sum_{i\in I_t} \mathbb{1}_{ij^*}$, one might observe that $p_t$ often deviates from target $p_b$, causing the company to either overspend and violate the budget constraint or underspend with suboptimal offered coupons.

According to Proposition \ref{prop:lambda}, the offered price is non-decreasing in  $\hat{\lambda}$. Thus, to mitigate the impact of a misspecified $\hat{\lambda}$, we make real-time adjustments to $\lambda$ to close the gap between $p_t$ and $p_b$ and make sure the company stays within its budget by the end of the day. Specifically, we apply the widely-used PID (Proportional-Integral-Derivative) control method (\citeNP{johnson2005pid}) to dynamically adjust $\lambda$ in real time so that the algorithm's average discounted price is adjusted towards the target $p_b$. The following equation describes the PID model's control signal $u(t)$. 
\begin{equation*}
\label{PID}
\begin{aligned}
u(t) = K_p e(t) + K_i \int_{t-T}^{t} e(\tau) d\tau + K_d \frac{d}{dt}e(t) 
\end{aligned}
\end{equation*}
The first term represents the proportional control, where $e(t) = p_b - p_t$ is the difference between the real-time average price and the target $p_b$ at time $t$. In the second term, $\int_{t-T}^{t} e(\tau) d\tau$ corresponds to the accumulation of errors from $t-T$ to $t$. In practice, it is discretized and calculated as $\sum_{\tau=t-T}^t e(\tau) \Delta t$. In the third term, $\frac{d}{dt}e(t)$ represents the future trend of $e(t)$ and is calculated as $(e(t)-e(t-\Delta t))/\Delta t$. PID control aims to minimize the discrepancy between the system's actual output and the target $p_b$ by adjusting the system's input, $\lambda$. Meanwhile, $K_p, K_i, K_d$ are the weights for the proportional, integral, and derivative terms, respectively. In practice, these weights are manually adjusted to ensure effective control and may vary from city to city. Let the $\lambda$ value at time $t$ be denoted as $\lambda(t)$. The real-time, regulated $\lambda(t)$ is expressed as:
\begin{equation*}
\begin{aligned}
\lambda(t_0) &= \hat{\lambda},\\
\lambda(t+\Delta t) &=\lambda(t) + u(t)\,.
\end{aligned}
\end{equation*}
Every day at 6 A.M., the initial value of $\lambda(t_0)$ is set to $\hat{\lambda}$, which is estimated using historical arrival data through Equation \eqref{eq:lag_lp}. Subsequently, at each time step, $\lambda(t)$ is updated by adding the control signal $u(t)$. This adjustment allows the induced coupon value to more closely adhere to the target budget limit.

\begin{figure}[t]  
\centering 
\hspace{0.1cm}
\includegraphics[width=\textwidth]{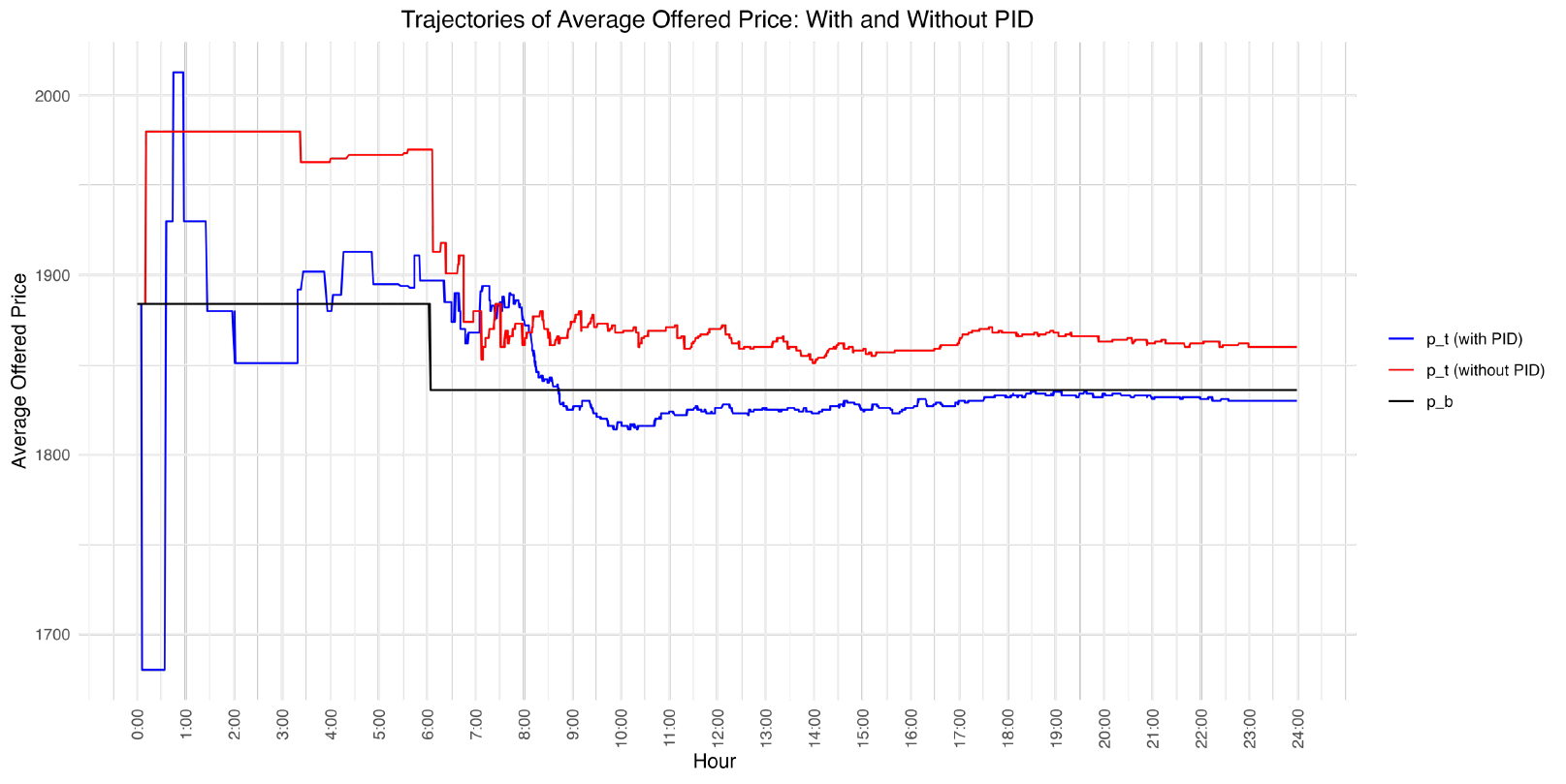} 
\caption{Trajectory of Average Offered Price $p_t$ throughout a Day} 
\label{Fig:PID_experiment} 
\end{figure}

Figure \ref{Fig:PID_experiment} illustrates the effectiveness of PID control in a real-life example. The blue and red trajectories represent the real-time average $p_t$ values with and without PID control, respectively. Every day, Meituan's system generates a new budget target $p_b$ at around 6 AM, and at the same time, the PID algorithm activates its control mechanism, as shown by the blue trajectory. Across the day, the controlled $p_t$ stays close to $p_b$, with a mere 1\% discrepancy on average. In contrast, when the firm does not update the estimated $\hat{\lambda}$ using PID, the resulting coupon values lead to a 3\% overshooting of the average offered price. This results in a substantial amount of unused budget and potentially a lower-than-optimal conversion rate.

\begin{figure}[t]  
\centering 
\includegraphics[width=1.0\textwidth]{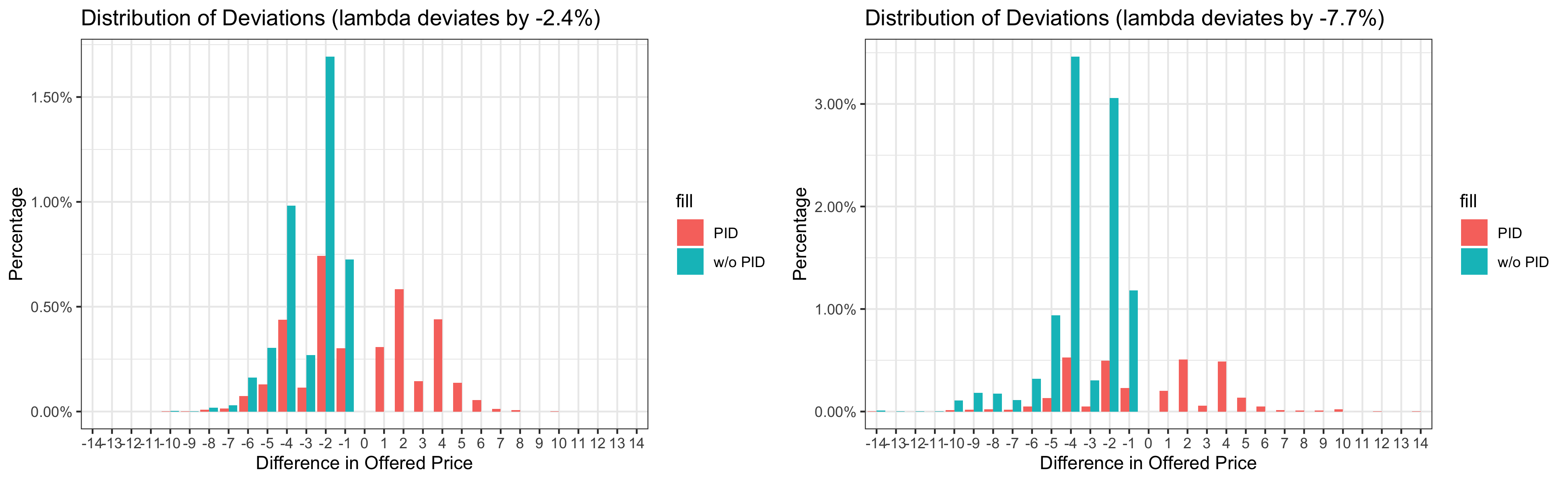} 
\caption{Distribution of Coupon Difference under $\lambda$ Misspecifications and PID Controls} 
\label{Fig:Dist_Diff_lambda_PID} 
\end{figure}

Finally, we investigate the resulting offered prices, budget targets, and the objective values in cases where  $\lambda$ is misspecified or adjusted by PID. To achieve this, we perform simulations drawing on historical data from two major cities where Meituan operates, which are referred to as City E and City F to maintain confidentiality. First, we obtain both the ground truth $\lambda^*$ and the corresponding optimal coupon value $j^*$ by solving Formulation \eqref{eq:basic_formulation}, with full knowledge of customer arrivals $I$. Meituan reports that the gap between the estimated $\hat{\lambda}$ and the ground truth optimal $\lambda^*$ is typically around 5\%. In our simulation, we perturb $\lambda^*$ by -2.4\% and -7.7\%, and assume these values as the $\hat{\lambda}$ used at the beginning of each day. These differences represent normal days and difficult-to-predict days, respectively. Under each scenario, we compare the offered prices for each individual customer between the model using the misspecified $\lambda$ and the ground truth oracle. The results are presented in Figure \ref{Fig:Dist_Diff_lambda_PID} and Table \ref{tab:lambda_pid}.

Our simulation results show that, in all four scenarios (high/low $\lambda$ deviation and with/without PID), the models offer the optimal prices to over 90\% of the customers even when $\lambda$ is misspecified. This suggests that Proposition \ref{prop_I_gap} holds to some extent in practice and that it is viable to use the estimated $\hat{\lambda}$ in the LDM approach. In Figure \ref{Fig:Dist_Diff_lambda_PID}, we display the price distribution for the remaining customers whose offered prices deviate from the optimal ones. It is clear that PID significantly reduces the proportion of customers with suboptimal prices, especially when $\hat{\lambda}$ has relatively high deviations. Moreover, Figure \ref{Fig:Dist_Diff_lambda_PID} shows that PID also reduces the magnitude of the biases in the offered price distributions. As a result, both the objective values and the budget target $p_b$ are much closer to the ground truth as indicated by Table \ref{tab:lambda_pid}.

\begin{table}[ht]
\centering
\small
\caption{Offline Experiments under $\lambda$ Misspecifications and PID Controls}
\begin{tabular}{lcccccc}
\hline
\multirow{2}{*}{\textbf{City}} &\textbf{Num. of}&\multirow{2}{*}{$\lambda$ \textbf{Deviation}}&\multirow{2}{*}{\textbf{Methods}} & \textbf{\% Users w/ Deviated}  & \textbf{Objective}& $p_b$ \\
&\textbf{Users}&&&\textbf{Offered Price}&\textbf{Deviation}&\textbf{Deviation}\\
\hline
City E &487,351&-2.4\%&without PID         &  4.18\%      &  +1.06\%       & -0.60\%\\
City E  &487,351&-2.4\% &with PID         &     3.52\%   &   +0.05\% &   -0.04\% \\
City F &544,109&-7.7\%&without PID         & 9.86\%       &  +2.66\%  &  -2.58\%  \\
City F  &544,109&-7.7\%&with PID      &    3.06\%    &      +0.04\% & -0.04\% \\
\hline
\end{tabular}\label{tab:lambda_pid}
\end{table}

\section{Experiments}\label{sec:experiments}
In this section, we test the effectiveness of our LDM-based coupon allocation against other benchmarks, such as the Offline IP approach discussed in Section \ref{sec:offline_ip}, using both comprehensive observational data and large online experiments. Through offline and online experiments, we assess the performance of LDM across various metrics, including computation time, coupon value distribution, its ability to stimulate customer demand, and ultimately, its financial implications.

\subsection{Summary of Coupon Allocation Strategies}\label{sec:algo_summary}
In the following sections, we compare the effectiveness of our proposed framework against several coupon pricing rules used by Meituan at different stages through offline and online experiments. To this end, we start by introducing the coupon allocation strategies used in the experiment. 
\begin{itemize}
\item \textbf{Manual}: In the Manual pricing strategy, the pricing manager first segments users into different groups. These groups include high-frequency, medium-frequency, low-frequency, churned, and new customers. The manager manually sets coupon values for each group to optimize customer CVR based on their experience. In addition, the pricing manager adjusts the pricing strategy manually so the end-of-the-day average price always coincides with the $p_b$.

\item \textbf{Random}: The Random pricing strategy allocates all coupon values in the ladder to users with equal probabilities. This approach does not incorporate any specific targeting or selection criteria. Such a random price exploration group is necessary to ensure that the CVR forecast model has enough variety in the training data to ensure identifiability.

\item \textbf{Offline-IP}: Offline-IP is an offline coupon allocation method based on the integer program characterized in Formulation \eqref{eq:offline_ip}. This method calculates the optimal coupon for all customers one day in advance and stores it in the database. When a customer arrives, the system queries the coupon value and immediately presents it to the customer. However, solving the optimal coupon value at the individual customer level is computationally infeasible for the Offline-IP approach. Thus, customers are first grouped into different groups according to their similarities. The integer programming then determines the optimal coupon value for each group.
    
\item \textbf{LDM}: In the LDM approach, the coupon value is determined in real time for each arriving customer based on Equations \eqref{eq:lag_lp}-\eqref{eq:optimal_j}. Notably, LDM employs the unadjusted CVR, $\tilde{q}_{ij}$, as input for these equations.

\item \textbf{LDM-IR}: Similar to LDM, LDM-IR also relies on Equations \eqref{eq:lag_lp}-\eqref{eq:optimal_j} to determine the coupon value. The key distinction is that LDM-IR utilizes the isotonic regression adjusted CVR $\hat{q}_{ij}$ as input, which ensures that the predicted CVR is non-decreasing concerning the coupon value.

\end{itemize}

\subsection{Offline Analysis of Simulation Results}\label{sec:offline_exp}
Drawing on comprehensive observational data from Meituan, we conduct offline simulations to compare the effectiveness of Offline-IP that Meituan used to implement and our proposed LDM-IR approach. The comparison focuses on three dimensions: (1) computation time, (2) the optimal objective value, and (3) the distribution of the offered prices. For this evaluation, we utilize estimated user CVR data from four cities, denoted as Cities C, D, E, and F, to ensure confidentiality. Both the offline-IP and LDM-IR algorithms are implemented and tested on a computing system equipped with a 2.6 GHz six-core Intel Core i7 processor.

First, we compare the computation time between the offline-IP and LDM-IR approaches. As mentioned in Section \ref{sec:offline_ip}, many cities that Meituan operates in have millions of users. Consequently, directly solving the Offline-IP, i.e., Formulation \eqref{eq:offline_ip}, would cause the computer to run out of memory. Following Meituan's current practice, we first cluster users into groups based on the similarity of their CVR and then solve Formulation \eqref{eq:offline_ip}, prescribing coupon value at the group level. We present the results in Tables \ref{tab:offline_1} and \ref{tab:offline_2}.

\begin{table}[htbp]
\centering 
\caption{Model Performance Comparison on Group Level for Two Large Cities.}
\small
\begin{tabular}{l l
c c c c c c}
\hline
\textbf{City}  & \textbf{Method} & \textbf{User Number} & \textbf{Group Number} & \textbf{Time/s} & \textbf{Objective Value} \\
\hline
\multirow{3}{*}{City C} & Offline-IP (by User) & 6,329,069 & - & OOM & NA\\ 
&Offline-IP (by Group)&-&2,029&83.42 & 583,676.17 \\
  &LDM-IR (by Group) &- & 2,029 &  0.28 & 583,674.40 \\
\hline
\multirow{3}{*}{City D}  &  Offline-IP (by User)&4,931,008 &-& OOM & NA \\
&Offline-IP (by Group)&-&1,773&61.95 & 391,943.73\\
  &LDM-IR (by Group)&-& 1,773 & 0.25 & 391,910.56 \\
\hline
\multicolumn{2}{l}{*OOM = Out of Memory}&&&&
\end{tabular}\label{tab:offline_1}
\end{table}

Table \ref{tab:offline_1} shows that the computer reports running out of memory when we directly attempt to solve Formulation \eqref{eq:offline_ip}. After grouping approximately 6 million users in City C into 2,029 groups, the LDM-IR approach is nearly 300 times faster than the Offline-IP method while achieving an almost identical objective value. Similarly, our simulation using offline data from City D yields comparable results.

\begin{table}[htbp]
\centering
\caption{Model Performance Comparison on Group Level for Two Medium Cities.}
\small
\begin{tabular}{l l
c c c c c c}
\hline
\textbf{City}  & \textbf{Method} & \textbf{User Number} & \textbf{Group Number} & \textbf{Time/s} & \textbf{Objective Value} \\
\hline
\multirow{3}{*}{City E}  &  Offline-IP (by User)&487,351 &-& OOM & NA \\
&Offline-IP (by Group)&-&1,508& 71.28 & 63,638.27\\
  &LDM-IR (By User) & 487,351& - & 30.18 & 70,902.23 \\
\hline
\multirow{3}{*}{City F}  &  Offline-IP (by User)&544,109 &-& OOM & NA \\
&Offline-IP (by Group)&-&1,115&48.79 & 86,759.31\\
  &LDM-IR (By User) & 544,109& - & 36.35 & 105,539.58 \\
\hline
\multicolumn{2}{l}{*OOM = Out of Memory}&&&&
\end{tabular}\label{tab:offline_2}
\end{table}

Table \ref{tab:offline_1} compares computation time for both the Offline-IP and LDM-IR methods when solving the group-level coupon allocation problem. In Table \ref{tab:offline_2}, we demonstrate the capability of LDM-IR to optimize coupon values at the individual level and report the corresponding running time. Although our LDM-IR approach recommends a single coupon value by solving Equation \eqref{eq:optimal_j} upon each new customer's arrival, resolving the entire optimization problem \eqref{eq:lag_lp} remains relevant. This is because the company must determine the $\lambda$ value offline in order to establish the $\hat{\lambda}$ for the following day. Efficiently solving the problem at the individual level results in a more accurate $\lambda$. Our findings indicate that the LDM-IR approach is not only faster than the group-level optimization provided by Offline-IP, but also achieves a higher objective value in comparison to its Offline-IP counterpart.

\begin{figure}[t]  
\centering 
\includegraphics[width=1.0\textwidth]{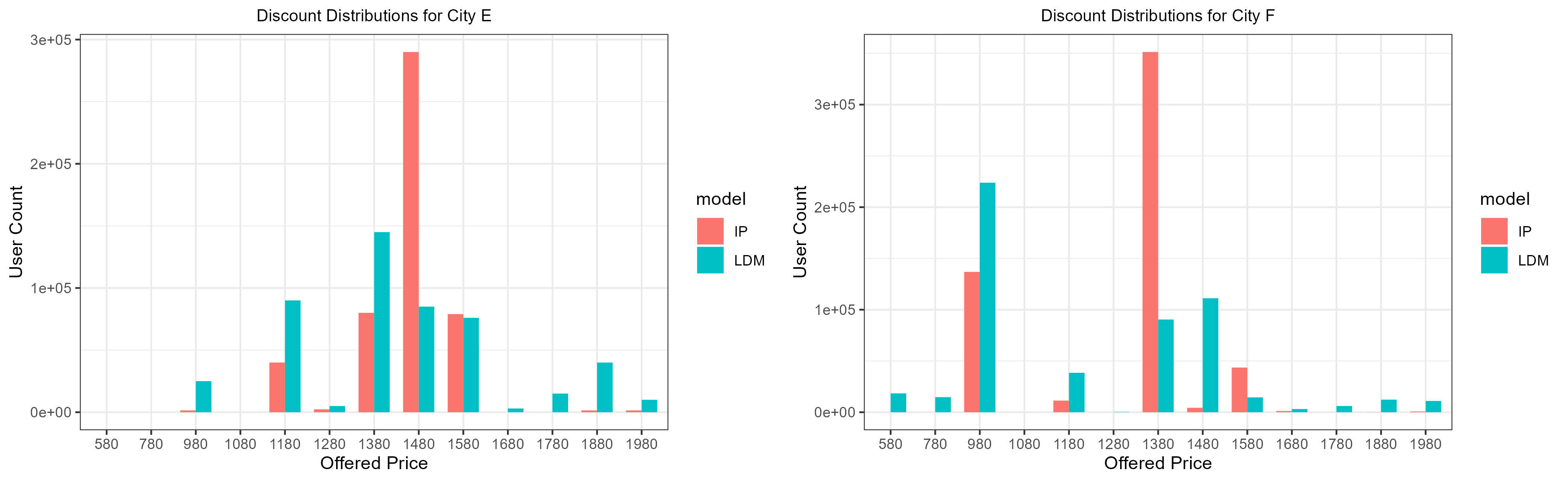} 
\caption{Comparison of Offline Offered Price Distribution between IP and LDM.} 
\label{Fig:simulation_distribution} 
\end{figure}

Finally, we explore the sources of the gain in the objective value reported in Table \ref{tab:offline_2} by examining the distributions of offered prices $p_j$ under Offline-IP and LDM-IR. These results are depicted in Figure \ref{Fig:simulation_distribution}. For the IP scenario, the prices are derived from Formulation \eqref{eq:offline_ip}, with $1,508$ and $1,115$ groups for the respective cities. For the LDM-IR scenario, the prices are determined from the relaxed Lagrangian Formulation \eqref{eq:lag_lp}. It is important to note that the price distribution under LDM-IR will be identical to that prescribed by Equation \eqref{eq:optimal_j}, assuming the estimated $\hat{\lambda}=\lambda^*$. Although this may not always be the case in reality, it still offers a glimpse into the recommended prices provided by the LDM-IR system. The most noticeable difference in the price distributions is that under LDM-IR, the offered prices are more dispersed compared to their Offline-IP counterparts. Since there is no need for customer grouping under LDM-IR, the system can better respond to the highly diverse individual features and prescribe coupon values with greater precision. This results in a more effective coupon allocation system than what is achieved with Offline-IP.

\subsection{Online Evaluation: Experimental Setup}\label{sec:online_exp}

In collaboration with Meituan Bike, we carry out large-scale experiments to evaluate the performance of our proposed LDM-based approach and compare its effectiveness to other coupon pricing strategies listed in Section \ref{sec:algo_summary}. 

In February 2024, we conducted two large-scale field experiments in two major cities in China, encompassing a random set of 3,867,738 users in total. Due to confidentiality agreements, these cities are referred to as City A and City B throughout the paper. Both experiments last for a month, and there are no major holidays or other concurrent promotions during the course of the experiment. Throughout the experiment, the company uses different pricing strategies to determine the coupon value for its monthly subscription plan to maximize the total revenue from subscribers. That is, $v_{ij}=\hat{q}_{ij}p_j$. In City A, the experiment took place from February 28th to March 28th, during which four coupon allocation methods were implemented: Manual, Random, LDM, and LDM-IR. Subsequently, in City B, we tested four coupon allocation methods: Manual, Random, Offline-IP, and LDM-IR from February 22nd to March 22nd. When a participating user logs into the app and opens the purchasing page, they are assigned to one of the groups via Meituan's A/B testing module and will remain in that group for the duration of the experiment. This module ensures that the traffic groups are orthogonal and well-balanced.

For each exposed user in a group, we record the price offered by the corresponding coupon pricing strategy and whether the user makes a purchase. Importantly, these different strategies affect only the coupon values for the monthly subscription plan. For other plans, such as weekly subscription or pay-per-use scheme, the coupon either has constant value or is not made available in the first place. Aside from the specific coupon value, the purchasing page (e.g., the page layout and order in which each plan is displayed) remains the same for users across different groups.


\subsection{Analysis of Online A/B Test}
We summarize the results of both experiments in Table \ref{tab:comparison}. Specifically, we compare the conversion rate, average offered price, and revenue per person of the proposed LDM-IR to other coupon delivery strategies.

\begin{table}[h]
    \centering
    \small
    \caption{A/B Results of Different Coupon Allocation Algorithms}
    \begin{tabular}{lccccc}
        \hline
        \multirow{2}{*}{\textbf{City}}&\multirow{2}{*}{\textbf{Exp. Group}}&\multirow{2}{*}{\textbf{Exposed Users}}&\multirow{2}{*}{\textbf{CVR}}&\multirow{2}{*}{\textbf{Avg.Price$^\dagger$}}&\multirow{2}{*}{\textbf{Avg.Price $\times$ CVR}}\\
         &  &  &  &  &  \\
        \hline
        \multirow{4}{*}{City A} & \textbf{Random} & 487,145 & 12.93\% & 17.82 & 2.31 \\
         & \textbf{Manual} & 778,916 & 13.59\% & 17.84 & 2.42  \\
         & \textbf{LDM} & 389,062 & 14.34\% & 17.24 & 2.47   \\
         & \textbf{LDM-IR} & 390,906 & 14.49\% & 17.51 & 2.54  \\
        \cline{1-6}
        \multirow{4}{*}{City B} & \textbf{Random} & 362,593 & 11.29\% & 17.78 & 2.01  \\
         & \textbf{Manual} & 174,483 & 12.10\% & 17.06 & 2.06  \\
         & \textbf{Offline-IP} & 640,964 & 12.56\% & 16.61 & 2.09 \\
         & \textbf{LDM-IR} & 644,669 & 12.93\% & 16.64 & 2.15 \\
        \hline
        \multicolumn{6}{l}{$^\dagger$ $p_b$ is set to be CNY 17.84 for all groups in City A and CNY 17.06 for all groups in City B.}
    \end{tabular}
    \label{tab:comparison}
\end{table}

First, we compare the CVR across various coupon pricing strategies. Although the primary objective is not to directly maximize CVR, we observe that for both City A and City B, LDM-IR leads to a higher CVR compared to Random and Manual approaches with $p<0.001$ from the $z$-test. To account for an increased probability of a Type I error due to multiple comparisons, we apply the Bonferroni correction (\citeNP{abdi2007bonferroni}) to reduce the family-wise error rate (FWER). The Bonferroni-corrected $p$-value still satisfies $p<0.001$. The observed CVR improvement over the Random and Manual approaches is not unexpected, as both lack an underlying optimization framework and distribute coupons either randomly or based on experience. Additionally, we observe that LDM-IR leads to a higher CVR compared to Offline-IP (with $z=-6.30$ and the corresponding $p<0.001$). This can be attributed to the fact that revenue maximization is performed at an individual level for LDM-IR, whereas it occurs at a group level for Offline-IP due to computational limitations. Interestingly, we do not observe a significant improvement in CVR when comparing LDM-IR to LDM ($p=0.057$), although the adjusted CVR, $\hat{q}_{ij}$, does yield a better PCOC score on the test data compared to the unadjusted $\tilde{q}_{ij}$. This might be attributed to the fact that LDM-IR, on average, prescribes a higher price to customers in the experiment, which we discuss in more detail in the following paragraph.

Another metric we analyze is the proximity of the average price prescribed by each pricing strategy to the predetermined price lower bound, $p_b$. In addition to maximizing conversion or revenue, it is crucial that coupon value is optimized to adhere to the agreed-upon marketing budget. With the Manual pricing strategy, the pricing manager is mindful of $p_b$ and consistently adjusts it so that the average offered price aligns with $p_b$ by the end of the day. This is why the average offered price is identical to $p_b$ for both cities under the Manual pricing strategy. On the other hand, the random pricing rule offers all coupon values with equal probability. As a result, the variation in average after-coupon prices across cities depends exclusively on differences in customer composition and their CVR rather than the specific $p_b$. This causes the average after-coupon price to potentially exhibit significant deviations from $p_b$, as City B illustrates. Additionally, we observe that although LDM-IR sets a higher price compared to LDM, it marginally falls below $p_b$ in both cities, which is consistent to the pattern illustrated in Figure \ref{Fig:PID_experiment}. 

Furthermore, we examine the revenue implications of various pricing rules. To align with the objective in both Offline-IP and LDM-based approaches, we first examine the average revenue generated exclusively from the monthly subscription plan, measured as Avg.Price$\times$ CVR. Using this metric, we observe from the experiment in City A a substantial and statistically significant revenue improvement with LDM-IR compared to Random, Manual, and LDM pricing strategies. The t-values for these comparisons are 107.57, 60.71, and 30.03, respectively, with corresponding $p<0.001$ in each case. We apply the Bonferroni correction to address the multiple comparison issues, and the results remain valid. Furthermore, the experimental results from City B not only reinforce the findings from City A, which suggest LDM-IR generates higher revenue than Random and Manual rules but also indicate that LDM-IR leads to increased revenue from the monthly subscription plan when compared to the Offline-IP case ($t=34.02$ and a $p<0.001$). The results from City B also suggest that replacing Offline-IP with LDM-IR could result in a 2.8\% increase in revenue per exposed user. This translates into approximately CNY 6 million (or USD 0.8 million) in monthly subscription revenue per 100 million exposed users.


\begin{table}[h]
    \centering
    \small
    \caption{A/B Results in CVR for Different Customer Groups in City B}
    \begin{tabular}{lccccc}
        \hline
        & \multicolumn{5}{c}{\multirow{2}{*}{\textbf{CVR for Different Customer Groups}}}\\
        &&&&&\\
        \cline{2-6}
          \multirow{2}{*}{\textbf{Exp. Group}}&  \multirow{2}{*}{\textbf{Churned Users}} &  \multirow{2}{*}{\textbf{New Users}} & \textbf{Low Freq.} & \textbf{Medium Freq.} &  \textbf{High Freq.} \\
         &&&\textbf{Users}&\textbf{Users}&\textbf{Users}\\
        \hline
          \textbf{Offline-IP} &5.64\% &	4.13\%&	8.75\%	&24.83\%	&36.69\% \\
          \textbf{LDM-IR}& 6.09\%&	4.58\%&	8.62\%&	24.87\%	&38.63\%\\
          \textbf{Relative Difference}&+8.03\%	&+10.72\%	&-1.48\%&	+0.16\%	&+5.28\%\\
        \hline
    \end{tabular}
    \label{tab:cvr_comparison}
\end{table}

Finally, we utilize the extensive customer-level feature data from City B provided by Meituan to identify which customer segments contribute the most to the CVR improvement when comparing the Offline-IP approach against our proposed LDM-IR. We adopt the same segments used in the Manuel couponing strategy, classifying customers into high-frequency, medium-frequency, low-frequency, churning, and new customers. User segments are based on days since the first ride and ride frequency in the last 28 days. New users have registered within the past 7 days, while those registered for more than 7 days are split into low, medium, and high frequency. Churned users haven't used the service in the last 28 days but have done so previously. The specific customer numbers for each segment are masked for confidentiality. 

We report the CVR differences in Table \ref{tab:cvr_comparison}. Specifically, we notice that the effect of our LDM-IR approach on conversion rate is not homogeneous across segments. Specifically, it leads to a significant increase in CVR for churned users, new users, and high-frequency users while having a minimal impact on medium-frequency users ($+0.16\%$) and a slight decrease for low-frequency users ($-1.48\%$). The difference in customer response is likely due to the customers in different segments having non-uniform responses to a more dispersed coupon value, as shown in Figure \ref{Fig:simulation_distribution}. Notably, we observe that LDM-IR is particularly effective in bringing back churned users ($+8.03\%$). This is important for the company, as churned customers will not generate value for the platform soon and may stop using the service indefinitely.

\section{Conclusion and Limitations}
In this paper, we propose a real-time online coupon allocation system for firms to stimulate consumer demand more effectively. At the heart of the system is a Lagrangian Dual-based algorithm that enables the determination of optimal coupon values for each user within a 50-millisecond window. In contrast to the traditional Offline-IP approach, which prescribes coupons in a batched fashion and relies heavily on predicted customer show-up probability instead of real-time consumer dynamics, our approach determines the coupon value one at a time when a customer arrives and is thus computationally light. Moreover, integrating real-time recalibrations through Isotonic Regression and Proportional–Integral–Derivative controllers enhances accuracy and reliability in response to irregular customer arrival patterns and data-generating processes.

We put the effectiveness of our proposed framework to the test through both field experiments and offline counterfactual studies. Our findings demonstrate that the proposed LDM not only achieves increased conversion rates but also contributes to higher revenues from the monthly subscription plan while adhering reasonably well to the predetermined marketing budget. Meituan has adopted our framework, resulting in a significant increase in revenue. As of May 2024, Meituan Bike has implemented the LDM-based system and distributed coupons to over 100 million users across more than 110 major cities in China. This implementation has led to a 0.7\% increase in revenue, which translates to an additional 8 million CNY in annual profit.

Our findings open avenues for further research into integrating machine learning and traditional tools in operations research to optimize marketing strategies in various other contexts. our paper has a few limitations. First, as our focus is to propose a readily usable and integrated, but not necessarily thoroughly optimized, coupon allocation framework, we do not explore alternative, potentially more advanced, algorithms for predicting CVR, nor do we investigate control mechanisms other than Proportional-Integral-Derivative (PID). Future work could investigate the performance improvement from employing such state-of-the-art algorithms in our framework. platforms may offer multiple SKUs (e.g., various subscription plans). The potential cannibalization effects among different SKUs could reduce the benefits of our proposed framework. Therefore, future research could explicitly consider the substitution effect and optimize coupon values for each SKU to further improve the effectiveness of the prescribed coupons.

\newpage
\bibliographystyle{chicago}
\bibliography{reference.bib}

\begin{thebibliography}{}

\bibitem[\protect\citeauthoryear{Abdi et~al.}{Abdi
  et~al.}{2007}]{abdi2007bonferroni}
Abdi, H. et~al. (2007).
\newblock Bonferroni and {\v{s}}id{\'a}k corrections for multiple comparisons.
\newblock {\em Encyclopedia of measurement and statistics\/}~{\em 3\/}(01),
  2007.

\bibitem[\protect\citeauthoryear{Albert and Goldenberg}{Albert and
  Goldenberg}{2022}]{albert2022commerce}
Albert, J. and D.~Goldenberg (2022).
\newblock E-commerce promotions personalization via online multiple-choice
  knapsack with uplift modeling.
\newblock In {\em Proceedings of the 31st ACM International Conference on
  Information \& Knowledge Management}, pp.\  2863--2872.

\bibitem[\protect\citeauthoryear{Anderson and Simester}{Anderson and
  Simester}{2001a}]{anderson2001sale}
Anderson, E.~T. and D.~I. Simester (2001a).
\newblock Are sale signs less effective when more products have them?
\newblock {\em Marketing science\/}~{\em 20\/}(2), 121--142.
\newblock a.

\bibitem[\protect\citeauthoryear{Anderson and Simester}{Anderson and
  Simester}{2001b}]{anderson2001price}
Anderson, E.~T. and D.~I. Simester (2001b).
\newblock Price discrimination as an adverse signal: Why an offer to spread
  payments may hurt demand.
\newblock {\em Marketing Science\/}~{\em 20\/}(3), 315--327.
\newblock b.

\bibitem[\protect\citeauthoryear{Barlow and Brunk}{Barlow and
  Brunk}{1972}]{barlow1972isotonic}
Barlow, R.~E. and H.~D. Brunk (1972).
\newblock The isotonic regression problem and its dual.
\newblock {\em Journal of the American Statistical Association\/}, 140--147.

\bibitem[\protect\citeauthoryear{Bawa and Shoemaker}{Bawa and
  Shoemaker}{1989}]{bawa1989analyzing}
Bawa, K. and R.~W. Shoemaker (1989).
\newblock Analyzing incremental sales from a direct mail coupon promotion.
\newblock {\em Journal of marketing\/}~{\em 53\/}(3), 66--78.

\bibitem[\protect\citeauthoryear{Chen and Chen}{Chen and
  Chen}{2015}]{chen2015recent}
Chen, M. and Z.-L. Chen (2015).
\newblock Recent developments in dynamic pricing research: multiple products,
  competition, and limited demand information.
\newblock {\em Production and Operations Management\/}~{\em 24\/}(5), 704--731.

\bibitem[\protect\citeauthoryear{Chen, Ma, Mandalapu, Nagarjan,
  Shanmugasundaram, Vassilvitskii, Vee, Yu, and Zien}{Chen
  et~al.}{2012}]{chen2012ad}
Chen, P., W.~Ma, S.~Mandalapu, C.~Nagarjan, J.~Shanmugasundaram,
  S.~Vassilvitskii, E.~Vee, M.~Yu, and J.~Zien (2012).
\newblock Ad serving using a compact allocation plan.
\newblock In {\em Proceedings of the 13th ACM Conference on Electronic
  Commerce}, pp.\  319--336.

\bibitem[\protect\citeauthoryear{Cheng, Liu, Dai, Zhang, Fang, and Zu}{Cheng
  et~al.}{2022}]{cheng2022adaptive}
Cheng, X., C.~Liu, L.~Dai, P.~Zhang, Z.~Fang, and Z.~Zu (2022).
\newblock An adaptive unified allocation framework for guaranteed display
  advertising.
\newblock In {\em Proceedings of the Fifteenth ACM International Conference on
  Web Search and Data Mining}, pp.\  132--140.

\bibitem[\protect\citeauthoryear{Chiu, Choi, Dai, Shen, and Zheng}{Chiu
  et~al.}{2018}]{chiu2018optimal}
Chiu, C.-H., T.-M. Choi, X.~Dai, B.~Shen, and J.-H. Zheng (2018).
\newblock Optimal advertising budget allocation in luxury fashion markets with
  social influences: A mean-variance analysis.
\newblock {\em Production and Operations Management\/}~{\em 27\/}(8),
  1611--1629.

\bibitem[\protect\citeauthoryear{Den~Boer}{Den~Boer}{2015}]{den2015dynamic}
Den~Boer, A.~V. (2015).
\newblock Dynamic pricing and learning: historical origins, current research,
  and new directions.
\newblock {\em Surveys in operations research and management science\/}~{\em
  20\/}(1), 1--18.

\bibitem[\protect\citeauthoryear{Dong, Kouvelis, and Tian}{Dong
  et~al.}{2009}]{dong2009dynamic}
Dong, L., P.~Kouvelis, and Z.~Tian (2009).
\newblock Dynamic pricing and inventory control of substitute products.
\newblock {\em Manufacturing \& Service Operations Management\/}~{\em 11\/}(2),
  317--339.

\bibitem[\protect\citeauthoryear{Fischer, Albers, Wagner, and Frie}{Fischer
  et~al.}{2011}]{fischer2011dynamic}
Fischer, M., S.~Albers, N.~Wagner, and M.~Frie (2011).
\newblock Dynamic marketing budget allocation across countries, products, and
  marketing activities.
\newblock {\em Marketing Science\/}~{\em 30}, 568--585.

\bibitem[\protect\citeauthoryear{Fong, Fang, and Luo}{Fong
  et~al.}{2015}]{fong2015geo}
Fong, N.~M., Z.~Fang, and X.~Luo (2015).
\newblock Geo-conquesting: Competitive locational targeting of mobile
  promotions.
\newblock {\em Journal of Marketing Research\/}~{\em 52\/}(5), 726--735.

\bibitem[\protect\citeauthoryear{Gopalakrishnan and Park}{Gopalakrishnan and
  Park}{2021}]{gopalakrishnan2021impact}
Gopalakrishnan, A. and Y.-H. Park (2021).
\newblock The impact of coupons on the visit-to-purchase funnel.
\newblock {\em Marketing Science\/}~{\em 40\/}(1), 48--61.

\bibitem[\protect\citeauthoryear{Gupta, Cotter, Pfeifer, Voevodski, Canini,
  Mangylov, Moczydlowski, and Van~Esbroeck}{Gupta
  et~al.}{2016}]{gupta2016monotonic}
Gupta, M., A.~Cotter, J.~Pfeifer, K.~Voevodski, K.~Canini, A.~Mangylov,
  W.~Moczydlowski, and A.~Van~Esbroeck (2016).
\newblock Monotonic calibrated interpolated look-up tables.
\newblock {\em The Journal of Machine Learning Research\/}~{\em 17\/}(1),
  3790--3836.

\bibitem[\protect\citeauthoryear{Hao, Peng, Ma, Wang, Jin, Hao, Chen, Bai, Xie,
  Xu, et~al.}{Hao et~al.}{2020}]{hao2020dynamic}
Hao, X., Z.~Peng, Y.~Ma, G.~Wang, J.~Jin, J.~Hao, S.~Chen, R.~Bai, M.~Xie,
  M.~Xu, et~al. (2020).
\newblock Dynamic knapsack optimization towards efficient multi-channel
  sequential advertising.
\newblock In {\em International Conference on Machine Learning}, pp.\
  4060--4070. PMLR.

\bibitem[\protect\citeauthoryear{Heilman, Nakamoto, and Rao}{Heilman
  et~al.}{2002}]{heilman2002pleasant}
Heilman, C.~M., K.~Nakamoto, and A.~G. Rao (2002).
\newblock Pleasant surprises: Consumer response to unexpected in-store coupons.
\newblock {\em Journal of Marketing Research\/}~{\em 39\/}(2), 242--252.

\bibitem[\protect\citeauthoryear{Holthausen~Jr and Assmus}{Holthausen~Jr and
  Assmus}{1982}]{holthausen1982advertising}
Holthausen~Jr, D.~M. and G.~Assmus (1982).
\newblock Advertising budget allocation under uncertainty.
\newblock {\em Management Science\/}~{\em 28\/}(5), 487--499.

\bibitem[\protect\citeauthoryear{Jiang, Zhang, Chen, Luo, Yang, Wang, Zhou,
  Zhu, and Gai}{Jiang et~al.}{2020}]{jiang2020dcaf}
Jiang, B., P.~Zhang, R.~Chen, X.~Luo, Y.~Yang, G.~Wang, G.~Zhou, X.~Zhu, and
  K.~Gai (2020).
\newblock Dcaf: A dynamic computation allocation framework for online serving
  system.
\newblock {\em arXiv preprint arXiv:2006.09684\/}.

\bibitem[\protect\citeauthoryear{Johnson and Moradi}{Johnson and
  Moradi}{2005}]{johnson2005pid}
Johnson, M.~A. and M.~H. Moradi (2005).
\newblock {\em PID control}.
\newblock Springer.

\bibitem[\protect\citeauthoryear{Kellerer, Pferschy, and Pisinger}{Kellerer
  et~al.}{2004}]{Kellerer2004}
Kellerer, H., U.~Pferschy, and D.~Pisinger (2004).
\newblock {\em The Multiple-Choice Knapsack Problem}, pp.\  317--347.
\newblock Berlin, Heidelberg: Springer Berlin Heidelberg.

\bibitem[\protect\citeauthoryear{Liu, Shen, Li, and Chen}{Liu
  et~al.}{2021}]{liu2021stimulating}
Liu, Q., Q.~Shen, Z.~Li, and S.~Chen (2021).
\newblock Stimulating consumption at low budget: evidence from a large-scale
  policy experiment amid the covid-19 pandemic.
\newblock {\em Management Science\/}~{\em 67\/}(12), 7291--7307.

\bibitem[\protect\citeauthoryear{Reimers and Xie}{Reimers and
  Xie}{2019}]{reimers2019coupons}
Reimers, I. and C.~Xie (2019).
\newblock Do coupons expand or cannibalize revenue? evidence from an e-market.
\newblock {\em Management Science\/}~{\em 65\/}(1), 286--300.

\bibitem[\protect\citeauthoryear{Sahni, Zou, and Chintagunta}{Sahni
  et~al.}{2017}]{sahni2017targeted}
Sahni, N.~S., D.~Zou, and P.~K. Chintagunta (2017).
\newblock Do targeted discount offers serve as advertising? evidence from 70
  field experiments.
\newblock {\em Management Science\/}~{\em 63\/}(8), 2688--2705.

\bibitem[\protect\citeauthoryear{Sethuraman and Mittelstaedt}{Sethuraman and
  Mittelstaedt}{1992}]{sethuraman1992coupons}
Sethuraman, R. and J.~Mittelstaedt (1992).
\newblock Coupons and private labels: A cross-category analysis of grocery
  products.
\newblock {\em Psychology \& Marketing\/}~{\em 9\/}(6), 487--500.

\bibitem[\protect\citeauthoryear{Smith, Seiler, and Aggarwal}{Smith
  et~al.}{2023}]{smith2023optimal}
Smith, A.~N., S.~Seiler, and I.~Aggarwal (2023).
\newblock Optimal price targeting.
\newblock {\em Marketing Science\/}~{\em 42\/}(3), 476--499.

\bibitem[\protect\citeauthoryear{Spiekermann, Rothensee, and
  Klafft}{Spiekermann et~al.}{2011}]{spiekermann2011street}
Spiekermann, S., M.~Rothensee, and M.~Klafft (2011).
\newblock Street marketing: how proximity and context drive coupon redemption.
\newblock {\em Journal of Consumer Marketing\/}~{\em 28\/}(4), 280--289.

\bibitem[\protect\citeauthoryear{Wu, Chen, Yang, Wang, Tan, Zhang, Xu, and
  Gai}{Wu et~al.}{2018}]{wu2018budget}
Wu, D., X.~Chen, X.~Yang, H.~Wang, Q.~Tan, X.~Zhang, J.~Xu, and K.~Gai (2018).
\newblock Budget constrained bidding by model-free reinforcement learning in
  display advertising.
\newblock In {\em Proceedings of the 27th ACM International Conference on
  Information and Knowledge Management}, pp.\  1443--1451.

\bibitem[\protect\citeauthoryear{Wu, Zhou, He, Han, Chen, and Zheng}{Wu
  et~al.}{2024}]{wu2024metasplit}
Wu, W., J.~Zhou, A.~He, S.~Han, J.~Chen, and B.~Zheng (2024).
\newblock Metasplit: Meta-split network for limited-stock product
  recommendation.
\newblock {\em arXiv preprint arXiv:2403.06747\/}.

\bibitem[\protect\citeauthoryear{Yan, Wang, Zhou, Lin, and Jiang}{Yan
  et~al.}{2023}]{yan2023end}
Yan, Z., S.~Wang, G.~Zhou, J.~Lin, and P.~Jiang (2023).
\newblock An end-to-end framework for marketing effectiveness optimization
  under budget constraint.
\newblock {\em arXiv preprint arXiv:2302.04477\/}.

\bibitem[\protect\citeauthoryear{Zhao, Hua, Yan, Zhang, Xu, and Yang}{Zhao
  et~al.}{2019}]{zhao2019unified}
Zhao, K., J.~Hua, L.~Yan, Q.~Zhang, H.~Xu, and C.~Yang (2019).
\newblock A unified framework for marketing budget allocation.
\newblock In {\em Proceedings of the 25th ACM SIGKDD International Conference
  on Knowledge Discovery \& Data Mining}, pp.\  1820--1830.

\end{thebibliography}

\newpage

\appendix
\setstretch{1.5}

\centerline{\Large\textbf{\textcolor{blue}{Appendix}}}
\centerline{\Large\textbf{Data-Driven Real-time Coupon Allocation in the Online Platform}}

\setcounter{equation}{0} \setcounter{lemma}{0}
\setcounter{proposition}{0} \setcounter{table}{0}
\setcounter{figure}{0}
\renewcommand{\theequation}{\thesection.\arabic{equation}}
\renewcommand{\thelemma}{\thesection.\arabic{lemma}}
\renewcommand{\theproposition}{\thesection.\arabic{proposition}}
\renewcommand{\thefigure}{\thesection.\arabic{figure}}
\renewcommand{\thetable}{\thesection.\arabic{table}}

\section{Illustration of Meituan's Coupon Realizations}
\label{app:snapshot}
\begin{figure}[H]  
\centering 
\includegraphics[width=1.0\textwidth]{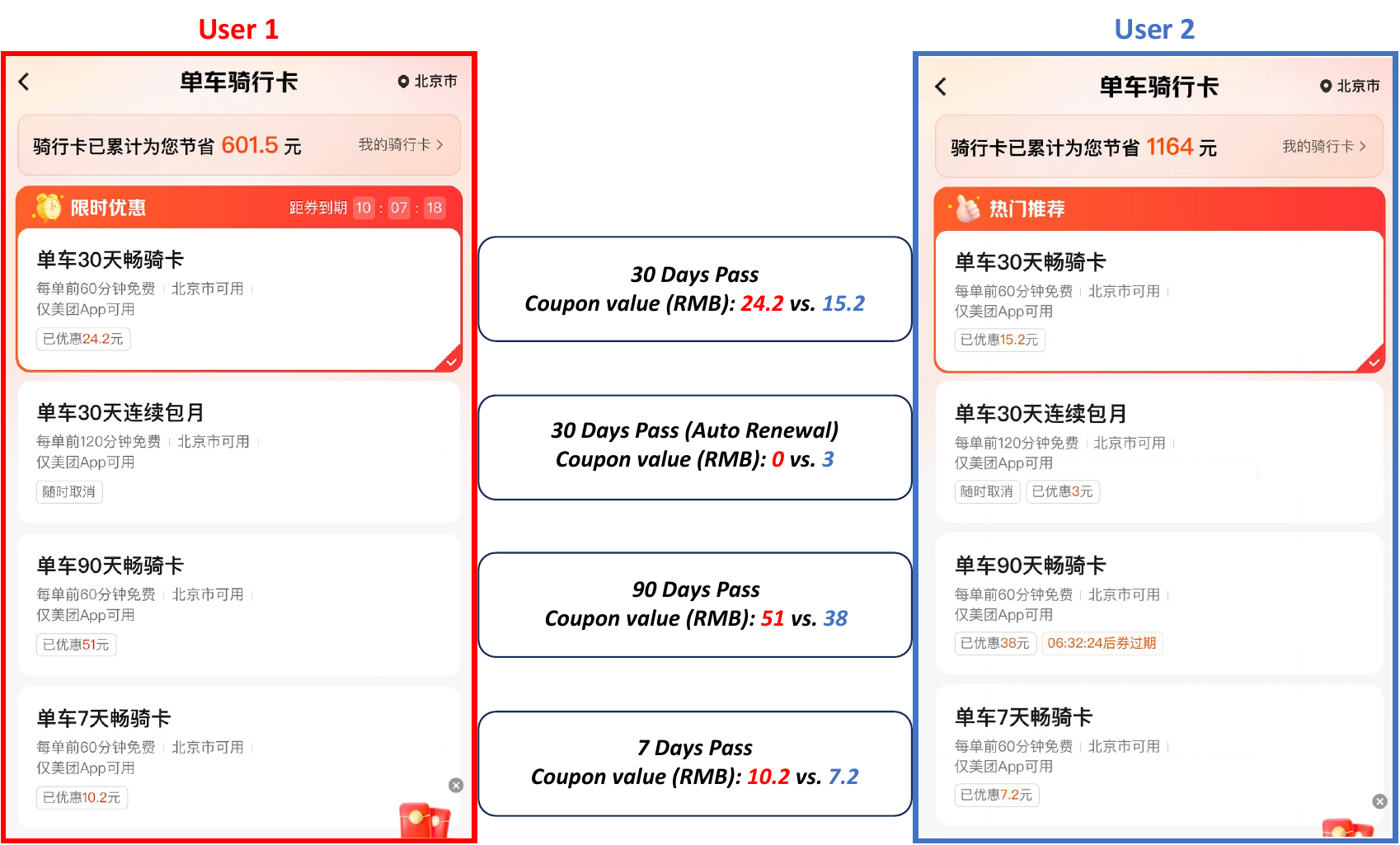} 
\caption{Snapshots of Different Coupon Values for Different Users.} 
\label{Fig:MT_Coupon} 
\end{figure}

\section{Numerical Setup for CVR Non-monotonicity}
\label{app:cvr}
We use a population of 100,000 customers with two features, $f_1$ and $f_2$. Feature 1 is the base utility term, which is drawn i.i.d from the standard normal distribution. Feature 2 is the price coefficient, which is drawn from the standard log-normal distribution. The price ladder is similar to the reality, starting from CNY 8 to CNY 16, with a CNY 2 stepsize. All prices are randomly allocated to all customers in the base setting. The extra campaign targets customers who are unwilling to make the purchase given the initial prices. For each campaign level, a random fraction of such target customers are selected and offered the updated prices of a low price (CNY 8). Customer $i$ will have a net utility in the following form:
\begin{align*}
    u_i = 10f_{1,i} - f_{2,i} \; p_i
\end{align*}
Customers will make the purchase if the utility $u_i > -6$. Note that there are no random noise terms in either the utility function or the purchasing decisions, so the only stochastic part of the simulation is the random allocation of prices. The company is assumed to be able to observe all the information, including customer features, offered prices, and their purchasing decisions. The XGBoost model is trained in R using the standard set of parameters that  $eta = 1$, $nthread = 2$, and $nrounds = 2$. We report that the monotonicity of the demand curve is violated when there is more than a 1\% increase in conversion rate for consecutive price levels.

\section{Proof of Propositions}
For ease of proof notations, we denote $V(i,j,\lambda)= v_{ij} - \lambda \hat{q}_{ij} (p_b - p_{j})$ from Equation \eqref{eq:optimal_j}.
\subsection{Proof of Proposition 1}
\label{app:proof}
The proof idea for proposition 1 is to show that when the sizes of two populations get big enough, we can always find two arriving customers, one in each population, that are similar enough. We start by defining such similarity. To this end, each customer can be represented by a point in an $N$ dimensional bounded feature space. Without loss of generality, we represent customer $i$ by $a_i = [a_{i,1},...,a_{i,N}] \in [0,1]^N$. We use the Chebyshev distance $d(a_i, a_{i'}) = max_n|a_{i,n} - a_{i',n}|$ to measure the distance between any two customers in the feature space.
\begin{definition}
Two populations $I_1$ and $I_2$ are called ``$\alpha_I$-similar'' if the following equation holds when customers are drawn randomly in each population.
\begin{equation*}
    \alpha_I  = \inf_{\alpha\in [0,1]}\Bigg\{\mathbb{P}_{x\in I_1}\Big(\inf_{y \in I_2} d(x,y) \leq \alpha\Big)\geq 1-\alpha \;\;\;\textnormal{and} \;\;\;\mathbb{P}_{y\in I_2}\Big(\inf_{x \in I_1} d(x,y) \leq \alpha\Big)\geq 1-\alpha\Bigg\}
\end{equation*}
\end{definition}
Intuitively speaking, $\alpha_I$ characterizes the closeness of random samples in each population. When $\alpha_I$ tends to zero almost surely, for any customer in $I_1$, we have an arbitrarily close customer in $I_2$. Providing that $I_1$ and $I_2$ have the same population size almost surely, we can argue that using either population will obtain the same Lagrangian multiplier $\lambda$. Leveraging this definition, we can rewrite Proposition 1 as the following Lemma.
\begin{lemma}
\label{lemma:prop1}
Two populations $I$ and $\hat{I}$ generated from $I_0$ are $\alpha_I$-similar, where $\alpha_I \to 0$ almost surely as $|I_0|\to \infty$.
\end{lemma}
Without loss of generality, we assume there are no common customers in $I$ and $\hat{I}$ because such customers will make the proof easier ($d(a_i,a_i) = 0$). Consider a union population $U = I\cup\hat{I}$, since both $I$ and $\hat{I}$ are constructed i.i.d from $I_0$, we get that $|I|,|\hat{I}| \sim Binomial(|U|,1/2)$,

For any given fixed level of $\alpha>0$, we can grid-partition the feature space $[0,1]^N$ into at most $(\lfloor1/\alpha\rfloor+1)^N$ cubes with side length $\alpha$. Define $S_i = \{x\in U:$ there are exactly $i$ other points in the same cube as $x\}$. We then know that $|S_i|\leq (i+1)(\lfloor1/\alpha\rfloor+1)^N$. Define $S_0 = \{x\in I: \inf_{y \in \hat{I}} d(x,y) > \alpha\}$ as the set of ``isolated points'' in $I$ that we cannot find any points in $\hat{I}$ in the same cube. We can bound $|S_0|$ in the following way, where $X_i \sim Binomial(|U|,1/2)$.
\begin{equation*}
    |S_0| \leq \sum_{i=0}^{|U|-1} |S_i|2^{X_i-i}\leq (\lfloor1/\alpha\rfloor+1)^N \sum_{i=0}^{|U|-1} (i+1)2^{X_i-i}
\end{equation*}
We are interested in the series $(i+1)2^{X_i-i}$. The mean is $(i+1)(3/4)^i$, and the variance is $(i+1)^2(5/8)^i$, both are convergent for the infinite sum series. Using Kolmogorov's two-series theorem, we get $\sum_{i=0}^\infty(i+1)2^{X_i-i}$ converges almost surely. As a result, $|S_0|/|U| \to 0$ as $|U| \to \infty$. Which recovers the result in Lemma \ref{lemma:prop1}.

\subsection{Proof of Proposition 2}
Given the multiplier $\hat{\lambda}$, consider $j^* = \mathrm{argmax}_{j\in J} V(i,j,\hat{\lambda})$ and a lower price level $p_{ij'}$ (if $j'$ does not exist, consider a higher price level $j'$ and the following analysis in the opposite direction). We know from Equation \eqref{eq:optimal_j} that $V(i,j^*,\hat{\lambda})\geq V(i,j',\hat{\lambda})$. If we have $\lambda'>\hat{\lambda}$, since $\hat{q}_{ij}$ is non-increasing in price $p_j$ from Section \ref{sec.isotonic}, we get $\hat{q}_{ij^*} \leq \hat{q}_{ij'}$. Denote $\lambda' = \hat{\lambda} + \epsilon$ and $\epsilon>0$, we can make the following transformations:
\begin{align*}
   V(i,j,\lambda')- V(i,j',\lambda') &=v_{ij^*} - \lambda' \hat{q}_{ij^*} (p_b - p_{j^*}) - v_{ij'} - \lambda' \hat{q}_{ij'} (p_b - p_{j'})\\
    &= V(i,j,\hat{\lambda})- V(i,j',\hat{\lambda})+ \epsilon(\hat{q}_{ij'}(p_b - p_{j'}) - \hat{q}_{ij^*}(p_b - p_{j^*}))\\
    &> 0 \;\;\; (\text{Since} \;\;\;\hat{q}_{ij'}\geq\hat{q}_{ij^*}, p_b - p_{j'}>p_b - p_{j^*})
\end{align*}
In other words, any lower price level $j'$ is not as good as the original price level $j$ when $\hat{\lambda}$ increases. Therefore, we conclude that the offered price level $p_{j^*}$ is non-decreasing in $\hat{\lambda}$ for all $i \in \hat{I}$.\\
We now compare with the ground-truth $\text{LP}_0$ with $\lambda^*$. Since $\lambda^*, \hat{\lambda}>0$, the budget constraint is nontrivial, it binds in the original LP relaxation from Formulation \eqref{eq:basic_formulation}. According to the complementary slackness in the LP duality theory, the budget constraint is violated if $\hat{\lambda}< \lambda^*$. Specifically, using the previous result,$\hat{\lambda}$ will lead to lower offered prices for some (at least one) customers, which further leads to the violation of the budget constraint in $\text{LP}_0$. On the other hand, if $\hat{\lambda} > \lambda^*$, we know the budget constraint is not binding in $\text{LP}_0$.

\subsection{Proof of Proposition 3}
First, we note that the results in proposition 3 are different in \shortciteNP{Kellerer2004}, where all the input parameters are assumed to be integers. In our case, the input parameters $\hat{q}_{ij}$ and $v_{ij}$ are generated from machine learning models with large dimensions of features and, therefore, unlikely to be integers. We can first write down the dual form for Formulation \eqref{eq:basic_formulation} as follows.
\begin{equation}
\label{dual_formulation}
\begin{aligned}
\max_{\lambda, \mu_i} \ &-\sum_{i \in \hat{I}}  \mu_i \\
s.t. \   &  \mu_i - V_{i,j,\lambda} \ge 0, \;\;\; \forall i \in \hat{I}, j \in J\\
          & \lambda \ge 0  \\
\end{aligned}
\end{equation}
It is clear that given a fixed level of $\lambda$, the optimal dual solution should set $\mu_i = max_j V_{i,j,\lambda}$. We know from complementary slackness in duality theory that if level $j'$ is not selected, meaning $\mu_i - V_{i,j',\lambda} > 0$, then the corresponding $x_{ij'}$ must be zero. Normally, if only one level of $j*$ is selected, then $x_{ij'} = 0 \forall j'\neq j$, and therefore $x_{ij^*}=1$, leading to integer solution for customer $i$. 
From assumption 2, Since $\hat{q}_{ij}$ and $v_{ij}$ are drawn from continuous distributions, we can obtain the following observation. 
\begin{equation*}
    \mathbf{P}(\hat{q}_{ij} = \hat{q}_{i'j'}) = 0, \; \mathbf{P}(v_{ij} = v_{i'j'} ) = 0, \;\;\; \forall (i,j) \neq (i',j') \in J
\end{equation*}
Intuitively, the probability of the event that any two customers have the exact same input parameter is zero. Therefore, given a fixed value of $\lambda_0$, the probability of the event that any two customers have the exact same $V_{i,j,\lambda_0}$ is also zero: $\mathbf{P}(V_{i,j,\lambda_0} = V_{i',j',\lambda_0}) = 0$. 

The reason we have fractional solutions in the LP relaxation is that there exists $(i,j,j')$ such that 
$\mu_i = V_{i,j,\lambda^*} = V_{i,j',\lambda^*}$. Here the extra degree of freedom from $\lambda$ allows the value of $ V_{i,j,\lambda^*}$ to be equal for different $j$ levels. Then since $\lambda^*$ is already fixed, for any other customers $i' \neq i \in \hat{I}$, 
we have $V_{i',j,\lambda^*}$ to be a random variable drawn from a continuous distribution
\begin{align*}
        \mathbf{P}(V_{i',j,\lambda^*} = V_{i',j',\lambda^*} | V_{i,j,\lambda^*} = V_{i,j',\lambda^*}) = 0, & \;\;\;\forall i' \neq i \\
    \mathbf{P}(V_{i,j,\lambda^*} = V_{i,j'',\lambda^*} | V_{i,j,\lambda^*} = V_{i,j',\lambda^*}) = 0, &\;\;\;\forall j''\neq j \
\end{align*}
In other words, there is at most one customer $i$ with fractional solutions, and for this customer $i$, there is at most one pair of fractional solutions.

\end{document}